\documentclass[journal]{IEEEtran}
\usepackage{blindtext}
\usepackage{graphicx}
\usepackage{graphicx,subfigure}
\usepackage{array}
\usepackage{url}
\usepackage{algorithmic}
\usepackage[cmex10]{amsmath}
\usepackage{ifpdf}
\usepackage{amssymb,amsmath}
\usepackage{empheq}
\usepackage{stfloats}   
\usepackage{cite}
\usepackage{algorithm}
\usepackage{algorithmic}

\usepackage[dvipsnames,usenames]{color}
\usepackage{prettyref}
\usepackage[normalem]{ulem}
\usepackage{amsthm}
\newtheorem{thm}{Theorem}
\newtheorem{cor}{Corollary}
\theoremstyle{remark}
\newtheorem{lem}{Lemma}
\newtheorem{propo}{Proposition}
\theoremstyle{remark}

\theoremstyle{remark}
\newtheorem{rmk}{Remark}
\usepackage{graphicx}
\usepackage{ifpdf}
\usepackage{cite}
\usepackage{enumerate}
\newcommand{\figref}[1]{\figurename~\ref{#1}}

\usepackage{array}
\usepackage{mdwmath}
\usepackage{mdwtab}
\usepackage{eqparbox}
\usepackage{bm}

\begin{document}

\title{Fundamental Limits of Training-Based Multiuser MIMO Systems}
 
\author{Xiaojun~Yuan,~\IEEEmembership{Senior Member,~IEEE,} Congmin~Fan,~\IEEEmembership{Student Member,~IEEE,} 
        and~Ying~Jun (Angela)~Zhang,~\IEEEmembership{Senior~Member,~IEEE}
\thanks{This work was supported by the National Natural Science Foundation of China (Project number 61471241, 61201262).}
\thanks{X. Yuan is with the School of Information Science and Technology, ShanghaiTech University, Shanghai, China, e-mail: yuanxj@shanghaitech.edu.cn.}
\thanks{C. Fan and Y.  Zhang are with the the Department of Information Engineering, The Chinese University of Hong Kong, Shatin, New Territories, Hong Kong. e-mail: yjzhang@ie.cuhk.edu.hk.}
}


\maketitle

\begin{abstract}
In this paper, we endeavour to seek a fundamental understanding of the potentials and limitations of training-based multiuser multiple-input multiple-output (MIMO) systems. In a multiuser MIMO system, users are geographically separated. So, the near-far effect plays an indispensable role in channel fading. The existing optimal training design for convenitional MIMO does not take the near-far effect into account, and thus is not applicable to a multiuser MIMO system. In this work, we use the majorization theory as a basic tool to study the tradeoff between the channel estimation quality and the information throughput. We establish tight upper and lower bounds of the throughput, and prove that the derived lower bound is asymptotically optimal for throughput maximization at high signal-to-noise ratio. Our analysis shows that the optimal training sequences for throughput maximization in a multiuser MIMO system are in general not orthogonal to each other. Futhermore, due to the near-far effect, the optimal training design for throughput maximization is to deactivate a portion of users with the weakest channels in transmission. These observations shed light on the practical design of training-based multiuser MIMO systems.
\end{abstract}


\IEEEpeerreviewmaketitle

\section{Introduction \label{sec:Intro}}
\IEEEPARstart{M}{ultiple} antenna (a.k.a. MIMO: multiple-input multiple-output) techniques have been extensively studied to improve the spectral efficiency of mobile communication systems, and are envisioned to be ubiquitously supported to accommodate the exponential growth of future wireless service demands. MIMO communications, however, requires the knowledge of the channel state information (CSI) at the transmitter for precoding and at the receiver for signal detection. A common approach is to allocate a certain amount of time and power resources to construct and transmit training signals for acquiring CSI. The impact of CSI acquisition on the overall performance of a MIMO system has been investigated under various performance measures, such as channel minimum mean-square error (MMSE), bit error rate (BER), and channel input output mutual information (MI) \cite{zheng2002communication, hassibi2003much, medard2000effect, Biguesh06, Minn06, Liu07, Coldrey07}. Among these measures, MI is advantageous in that it characterizes the fundamental tradeoff between achieving high-quality channel estimate and the information throughput. One one hand, to achieve a high-quality channel estimate, more time and power resources should be allocated for training, leaving the data transmission with little time and power. On the other hand, if too little resources are allocated to training, information throughput deteriorates due to channel mismatch. The exact throughtput characterization of the training-based MIMO system is difficult, but a tight MI lower bound was derived in \cite{zheng2002communication, hassibi2003much}. The authors in \cite{zheng2002communication} also discussed the tradeoff from the perspective of degrees of freedom (DoF) \cite{zheng2002communication}. Later, Coldrey et al. established the optimal tradeoff by assuming independent and identically distributed (i.i.d.) Rayleigh fading and exploiting the rotational invariance property of an i.i.d. Gaussian channel matrix \cite{Coldrey07}.

Recently, the incorporation of MIMO into multiuser cellular systems has attracted much research interest, especially in a  massive MIMO setup where users communicate with a base station (BS) equipped with a large-scale antenna array \cite{marzetta2010noncooperative, scalingup, ngo2011energy, hoydis2013massive}. A multiuser MIMO system can be treated as a virtual MIMO system without cooperation at the transmitter side. It exhibits some new features compared with conventional MIMO. First, geographically separated users in general experience significantly different large-scale fading caused by path loss and shadowing. That is, the near-far effect is indispensible in modelling a multiuser MIMO system. Second, the total transmission power of a MIMO system is usually constrained, while that of a multiuser MIMO system scales with the number of active users. Third, large antenna arrays may be deployed at BSs, focusing energy into ever-smaller spatial regions to bring huge improvements in system throughput and energy efficiency; but at the same time, large antenna arrays creates a lot more channel links than ever before, which imposes a heavy burden on the acquisition of CSI. The above features imply that the pilot design presented in \cite{Coldrey07} (under the Rayleigh fading assumption and a total transmission power constraint) is not necessarily good for a multiuser MIMO system. New insights and guidelines must be developed to better understand the tradeoff between the channel estimation quality and the information throughput in a multiuser MIMO system.

In this paper, we investigate the fundamental throughput limit of a training-based MIMO multiple-access system, where each transmission frame consists of a training phase for acquiring CSI and a data-transmission phase for information delivery. We assume that the system consists of an $N$-antenna BS and $K$ single-antenna user terminals, forming an $N$-by-$K$ virtual MIMO channel with coherence time $T$ (during which the channel state is assumed to be constant).  Each user suffers both large-scale fading caused by the near-far effect and small-scale fading caused by the multipath effect. As large-scale fading varies much slowly in magnitude of time than small-scale fading \cite{Rappaport}, we assume that the large-scale fading coefficients of users are known \emph{a priori} at the BS, while the small-scale fading coefficients of users are to be estimated using training sequences. Our target is to characterize and optimize the throughput of a training-based multiuser MIMO system over the parameters including the pilot symbols, the time allocation coefficient $\alpha$ (which specifies the fraction of the training phase in a transmission frame), the power allocation coefficient $\gamma_k$ of each user $k$, and the user number $K$, etc.

Due to the near-far effect, the distribution of the MIMO channel matrix is rotationally variant. As a result, the technique developed in \cite{Coldrey07} is not applicable to a MIMO multiuser system. Instead, we use the majorization theory \cite{Majortheory, Palomar2003} as a basic tool to derive upper and lower bounds of the system throughput. We show that the derived throughput lower bound is asymptotically optimal for throughput maximization in the high SNR regime. We also show that the upper and lower bounds are reasonably tight in various system settings. We note that the results in this paper is applicable to an arbitrary antenna and user configuration. Further, to establish a close link with massive MIMO, we use the random matrix theory to derive a closed-form expression of the system throughput as $N$, $K$, and $T$ scale at the same speed towards infinity, under the assumption of uniform large-scale fading (i.e., users are co-located).

An interesting finding of this work is that, in a training-based multiuser MIMO system, the optimal training length $\alpha T$ for throughput maximization is usually less than the number of active users $K$. This is in contrast with the case of conventional MIMO in which the optimality always occurs at $K = \alpha T$ (implying that each user has one separate time slot for channel estimation) \cite{Coldrey07}. To understand this fact, we first note that a MIMO multiuser system does not reduce to conventional MIMO even if uniform large-scale fading is assumed. The key difference is that the total transmission power of a MIMO multiuser system scales linearly with the number of active users, while that of a conventional MIMO system is usually limited by a fixed total power budget. Consequently, for a multiuser MIMO system, the scalability of the total transmission power shifts the optimality point from $K = \alpha T$ to $K \leq \alpha T$. Particularly, when the optimality occurs at $K < \alpha T$, there is not enough degrees of freedom to design orthogonal training sequences (with length $\alpha T$) for all the $K$ users. This is again different from the case of conventional MIMO in which the optimal training sequences are always orthogonal \cite{Coldrey07}.

The disparity between the optimal $\alpha T$ and $K$ is further enlarged by the near-far effect. In fact, for throughput maximization in the considered multiuser MIMO system, a portion of users with relatively poor channel quality should be kept silent in transmission. This is because, with the near-far effect, the channel qualities of users vary significantly from each other. The channels of far-off users are so weak that it will be a waste of resource for throughput enhancement if any time or power is allocated to these users. As such, a good strategy is to inactivate these far-off users in transmission, which translates to an enlarged gap between the optimal $\alpha T$ and $K$.

\subsection{Other Related Work}
Existing work related to the throughput analysis of multiuser MIMO systems includes \cite{Marzetta06, Kobayashi11, Chi11}. Specifically, Marzetta examined the training-throughput tradeoff in a multiuser MIMO broadcast channel by assuming orthogonal training sequences \cite{Marzetta06}. Kobayashi et. al studied the throughput optimisation of a multiuser MIMO system by taking into account the effect of channel estimation error and finite channel state feedback \cite{Kobayashi11}. Chi et al. considered the pilot sequence design for a multiuser MIMO OFDM system, and derived the optimal pilot design in the sense of minimizing the mean-square error (MSE) of the channel estimation \cite{Chi11}. Given the above work, the characterization of the optimal training-throughput tradeoff is still missing in the literature, which motivates the work presented in this paper.

\subsection{Organization}
The remainder of this paper is organized as follows. First, in \prettyref{sec:System-model}, we describe our system model and formulate the throughput optimization problem. Then, in Section \ref{Section III} and Section \ref{Section IV}, we establish upper and lower bounds of the system throughput under the assumption of arbitrary large-scale fading. Later in Section \ref{Section V}, we derive the optimal design of the  system parameters under the assumption of uniform large-scale fading. We conclude the paper in Section \ref{Section VI}.

\subsection{Notation}
 Bold upper-case letters denote a matrix and bold-lower case letters denote a column vector.
 For a matrix $\mathbf{H}$, the element of row $i$ and column $j$ is denoted as $H_{ij}$.  
  $\mathbf{I}_{n}$ denotes an $n\times n$ indentity matrix, where $n$ is an integer.
The superscripts $\left(\cdot\right)^{\rm T}$, $\left(\cdot\right)^{\dagger}$ stand for the transpose and Hermitian transpose, respectively. The operators $\left(\cdot\right)^{-1}$, $\left|\cdot\right|$, $\mbox{tr}\left(\cdot\right)$ represent the inverse,  the determinant, and the trace of a matrix, respectively.  We use  $\left\Vert \cdot\right\Vert $ to denote the  norm of a vector, and $\otimes$ to denote the Kronecker product.
The vector inequality $\mathbf{x} \succ \mathbf{y}$ means that $\mathbf{x}$ majorizes $\mathbf{y}$; log denotes logarithm with base 2.
$\left(x\right)^{+}=\max\left\{ 0,x\right\} $; ${\rm diag}\{a_1, \cdots, a_n\}$ represents the diagonal matrix with the $(i,i)$th element being $a_i$; $(\mathbf{A})_{\rm diag}$ represents the diagonal matrix obtained by setting the off-diagonal elements of $\mathbf{A}$ to zeros; $\text{vdiag}\{\mathbf{A}\}$ represents the vector specified by the diagonal of $\mathbf{A}$, with the $i$th entry of $\text{vdiag}\{\mathbf{A}\}$ being $A_{ii}$. For a square matrix $\mathbf{A}$, $\bm{\lambda}(\mathbf{A})$ represents the vector of the eigenvalues of $\mathbf{A}$ (counting multiplicity) arranged in a descending order.

\section{Preliminaries \label{sec:System-model}}
\subsection{System Model}

Consider a multiuser MIMO system, where $K$ single-antenna users deliver information to an $N$-antenna base station (BS).  Assume that both $K$ and $N$ are very large but finite. The channel is block-fading, i.e., the channel keeps invariant within coherence time $T$. The corresponding channel model for a frame of $T$ symbols is given by
\begin{subequations}\label{Channel}
\begin{equation}
\mathbf{Y}=\sum_{k=1}^K d_k\mathbf{h}_k \mathbf{x}_k^{\rm T}+\mathbf{W} ,  \label{Whole_channel_MAC} 
\end{equation}
or equivalently
\begin{equation}
\mathbf{Y=HDX+W}, \label{eq:Whole_channel_model}
\end{equation}
\end{subequations}
where $\mathbf{Y}\!\in\!\mathbb{C}^{N\times T}$ represents the received signal matrix at the BS, and $\mathbf{X}\!=\!\left[\mathbf{x}_{1},{\cdots},\mathbf{x}_{K} \right]^{\rm T}\!\in\!\mathbb{C}^{K\times T}$ is the transmit signal matrix with the  $k$-th row $\mathbf{x}_k^{\rm T}$ being the signal of user $k$,  and $\mathbf{H}\in\mathbb{C}^{N\times K}$ is the small-scale channel fading matrix with the $(i,j)$-th element $H_{ij}$ connecting the $j$-th user to the $i$-th receive antenna of the BS, $\mathbf{D} = \text{diag}\{d_1, d_2,\cdots, d_K\}$ with each $d_k$ being a nonnegative large-scale fading coefficient of user $k$, and $\mathbf{W}\!\in\!\mathbb{C}^{N\times T}$ is the white Gaussian noise matrix with the power of each element given by $N_{0}$. Without loss of generality, we always assume that $\{d_k\}$ are arranged in a descending order, i.e., $d_1\geq d_2\geq\cdots\geq d_K$. The power constraint of each user $k$ is given by
\begin{equation}
\frac{1}{T}  \lVert \mathbf{x}_k \rVert ^2  \leq  P_{0},~~ k\in \mathcal{I}_K\!\triangleq\! \{ 1,2,\cdots,K \}.  \label{eq:pow_cons_1}
\end{equation}

The large-scale fading, due to signal propagation over large distances and shadowing from obstacles in the propagation path, usually varies relatively slowly. We assume that the BS antennas are geographicaly co-located and therefore, it suffices to use one coefficient $d_k$ to characterize the large-scale fading between user $k$ and the BS antennas. We further assume that $d_1, \cdots, d_K$ are known {\it{a priori}} at the BS, as the BS is able to acquire an accurate estimate of each $d_k$ based on historical data. The small-scale fading is caused by multipath propagation. We assume that the channel matrix $\mathbf{H}$ for small-scale fading follows independent Rayleigh fading, with each $H_{ij}$ independently drawn from $\mathcal{CN}(0,1)$. Note that $\mathbf{H}$ varies from frame to frame, and needs to be estimated at the BS based on the received data in each transmission frame.

When users are co-located, we have $d_1 = \cdots = d_K = d$, or equivalently, $\mathbf{D} = d\mathbf{I}_K$, for a certain coefficient $d$. In this case, the system in (\ref{Channel}) is very similar to a conventional MIMO system. The only difference resides in the power constraint: In the considered multiuser MIMO system, the total transmission power is linear in the number of users, while in a conventional MIMO system, the total power constraint is usually invariant to the number of transmit antennas (i.e., the number of users in our model). In general, the system in (\ref{Channel}) allows the existence of the near-far effect, i.e., the users are geographically separated in a random manner, thereby resulting in different values of $\{d_k\}$.

The transmission protocol for (\ref{Channel}) is described as follows. We adopt a training-based scheme in which each transmission frame consists of  two phases. In the first phase (referred to as the training phase), pilot symbols known to the receiver are transmitted, based on which the channel matrix $\mathbf{H}$ is estimated. In the second phase (referred to as the data transmission phase), data are transmitted and detected based on the estimated channel. The details of these two phases are described below.

\subsection{Training Phase \label{sub:Uplink-training}} 
Without loss of generality, we assume that $\alpha T$ channel uses are assigned to the training phase, where $\alpha \!\in \!(0,1)$ is a coefficient to be optimized.
From \prettyref{eq:Whole_channel_model}, the channel model for  the training phase is 
\begin{align}
\mathbf{Y}_{\rm p} & =\mathbf{HD}\mathbf{X}_{\rm p}+\mathbf{W}_{\rm p},
\label{eq:Y_p}
\end{align}
where $\mathbf{X}_{\rm p}\!\in\! \mathbb{C}^{K\times \alpha T}$ is the pilot symbol matrix with $\mathbf{x}_{{\rm p},k}$ being the transpose of the  $k$-th row, and $\mathbf{W}_{\rm p}\in \mathbb{C}^{N\times \alpha T}$ is the corresponding AWGN.
The power of user $k$ in the training phase is given by
\begin{equation}
\frac{1}{\alpha T}  \lVert \mathbf{x}_{{\rm p},k}\rVert^2
=  \gamma_k P_0 , ~~k\in \mathcal{I}_K ,   \label{eq:pow_cons_2}
\end{equation} 
where $\mathbf{x}_{{\rm p}, k}^{\rm T}$ is the $k$th row of $\mathbf{X}_{\rm p}$, and $\gamma_k$ is a power allocation coefficient of user $k$ for the training phase.

The base stations use $\mathbf{X}_{\rm p}$ and $\mathbf{Y}_{\rm p}$ to generate an estimate of the channel $\mathbf{H}$, denoted as $\mathbf{\widehat{H}}=f(\mathbf{X}_{\rm p},\mathbf{Y}_{\rm p})$.
Particularly, the minimum mean-square error (MMSE) estimate of $\mathbf{H}$ is given by
\begin{equation}
\widehat{\mathbf{H}} = \mathbf{Y}_{\rm p}\left(\mathbf{X}_{\rm p}^{\dagger}\mathbf{D}^{2}\mathbf{X}_{\rm p}+N_{0}\mathbf{I}_{\alpha T}\right)^{-1}\mathbf{X}_{\rm p}^{\dagger}\mathbf{D}.
\label{eq:estimateH}
\end{equation}
The corresponding MMSE matrix is given by
\begin{eqnarray}
\mathbf{R}_{\rm MMSE}  &=& \mathbb{E}\left[\mbox{vec}(\mathbf{H}-\widehat{\mathbf{H}})\left(\mbox{vec}( \mathbf{H}-\widehat{\mathbf{H}})\right)^\dagger \right] \nonumber\\
&=& \mathbf{I}_N \otimes \mathbf{M}_{\widehat{\mathbf{H}}},
\end{eqnarray}where vec$(\mathbf{H}{-}\mathbf{\widehat{H}})$ is the transpose of the row vector obtained by sequentially stacking the rows of $(\mathbf{H}{-}\mathbf{\widehat{H}})$, the expectation $\mathbb{E}$ is taken over  $\mathbf{H}$ and $\mathbf{W}$, and 
\begin{align}\label{eq:MMSE_matrix} 
\mathbf{M}_{\widehat{\mathbf{H}}} = \mathbf{I}_{K}-\mathbf{D}\mathbf{X}_{\rm p}\left(\mathbf{X}_{\rm p}^{\dagger}\mathbf{D}^2\mathbf{X}_{\rm p}+N_{0}\mathbf{I}_{\alpha T}\right)^{-1}\mathbf{X}_{\rm p}^{\dagger}\mathbf{D}.
\end{align} 

\subsection{Data Transmission Phase}
In the data transmission phase, the users transmit data and the base stations carry out coherent
detection based on the channel estimate obtained in the training phase.
The channel model is written as
\begin{subequations}\label{eq:Y_data_V}
\begin{align}
&\mathbf{Y}_{\rm d}  = \widehat{\mathbf{H}}\mathbf{D}\mathbf{X}_{\rm d} + \mathbf{V},\label{eq:Y_d}  \\
\intertext{where}  
&\mathbf{V}\triangleq (\mathbf{H-\widehat{H}})\mathbf{D}\mathbf{X}_{\rm d}+\mathbf{W}_{\rm d},  \label{eq:V}
\end{align}
\end{subequations}
and $\mathbf{X}_{\rm d} \in\mathbb{C}^{K\times\left(1-\alpha\right)T}$
is a zero-mean data matrix, and $\mathbf{W}_{\rm d}$ is the corresponding AWGN.
The power consumption at user $k$ is expressed as
\begin{equation} 
\frac{1}{\left(1-\alpha\right)T} \lVert \mathbf{x}_{{\rm d},k} \rVert ^2 = \gamma^\prime_k P_0,    \label{eq:pow_cons_3}
\end{equation}
where $\mathbf{x}^{\rm T}_{{\rm d},k}$ is the $k$-th row of $\mathbf{X}_{\rm d}$, and $\gamma^\prime_k$ is a coefficient of user $k$.
With \eqref{eq:pow_cons_2} and \eqref{eq:pow_cons_3}, the power constraint in \eqref{eq:pow_cons_1} can be equivalently expressed as
\begin{equation}
\alpha \gamma_k + (1-\alpha)\gamma'_k \leq 1.
\end{equation}
The covariance matrices of $\mathbf{X}_{\rm d}$ and $\mathbf{V}$ are respectively given by 
\begin{subequations}\label{eq:RXd_RV}
\begin{align}
\mathbf{R}_{\mathbf{X}_{\rm d}} &	\triangleq \tfrac{1}{(1-\alpha)T} \mathbb{E}\left[ \mathbf{X}_{\rm d} \mathbf{X}_{\rm d}^\dagger \right] = \text{diag}\{\gamma'_1 P_0, \cdots, \gamma'_K P_0\} \label{eq:RXd_RV_a}\\
\mathbf{R}_{\mathbf{V}} 		&\triangleq \tfrac{1}{\left( 1-\alpha\right) T}\mathbb{E}\left[ \mathbf{V} \mathbf{V}^\dagger \right]\label{eq:RXd_RV_b}\\
&= \tfrac{1}{(1-\alpha)T}\mathbb{E}\left[(\mathbf{H}{-}\mathbf{\widehat{H}})\mathbf{D}\mathbf{R}_{\mathbf{X}_{\rm d}}\mathbf{D}(\mathbf{H}{-}\mathbf{\widehat{H}})^\dagger\right]+N_0\mathbf{I}_N\label{eq:RXd_RV_c}\\
&= \sigma_v^2 \mathbf{I}_N,
\end{align}
\end{subequations}
where the equivalent noise power is given by
\begin{equation}\label{eq:equivalent_noise_power}
\sigma_v^2 =  \mbox{tr} \left(\mathbf{M}_{\widehat{\mathbf{H}}}\mathbf{D}^2\mathbf{R}_{\mathbf{X}_{\rm d}}\right)+N_0.
\end{equation}
In the above, \eqref{eq:RXd_RV_a} follows from the fact that the user signals are independent of each other; \eqref{eq:RXd_RV_b} follows by noting $\mathbf{V}$ in \eqref{eq:V}, and \eqref{eq:RXd_RV_c} by (\ref{eq:MMSE_matrix}) and noting that the rows of 
$\mathbf{H} - \mathbf{\widehat{H}}$ are independent of each other.

Recall the signal model in \eqref{eq:Y_data_V}. The interference-plus-noise term $\mathbf{V}$ is in general correlated with the signal $\widehat{\mathbf{H}}\mathbf{DX}_{\rm d}$, which complicates the analysis.
However, it is known that the ``worst-case''  noise for the additive channel in \eqref{eq:Y_d} follows an independent Gaussian distribution \cite{medard2000effect}.
That is, the instantaneous achievable rate over the channel \eqref{eq:Y_d} is lower bounded by 
\begin{align}
I\left(\mathbf{X}_{\rm d};\mathbf{Y}_{\rm d} \right | \mathbf{\widehat{H}}) 
&=  \mbox{log} \left|\mathbf{I}_{N}+\mathbf{R}_{\mathbf{V}}^{-1}\widehat{\mathbf{H}}\mathbf{D}\mathbf{R}_{\mathbf{X}_{\rm d}}\mathbf{D}\widehat{\mathbf{H}}^{\dagger}\right| \notag \\
& = \mbox{log} \left|\mathbf{I}_{N}+\tfrac{1}{\sigma_{v}^2}\widehat{\mathbf{H}}\mathbf{D}\mathbf{R}_{\mathbf{X}_{\rm d}}\mathbf{D}\widehat{\mathbf{H}}^{\dagger}\right| , \label{eq:lower_bound}
\end{align}
where  $I\left(\mathbf{X}_{\rm d};\mathbf{Y}_{\rm d} \right | \mathbf{\widehat{H}}) $ is the conditional mutual information between $\mathbf{X}_{\rm d}$ and $\mathbf{Y}_{\rm d}$ calculated by assuming that the elements of the $k$th row of $\mathbf{X}_{\rm d}$ are independently drawn from $\mathcal{CN}(0,\gamma'_k P_0 )$ for $k = 1, \cdots, K$,  and those of  $\mathbf{V}$ are independently drawn from $\mathcal{CN}(0,\sigma_v^2 )$. Then, by considering the two-phase protocol and averaging over the channel fading, we obtain an achievable throughput of the system given by
\begin{equation}
R =\left(1{-}\alpha \right) {\mathbb{E}}\left[ \log\left|\mathbf{I}_{N}{+}\tfrac{1}{\sigma_{v}^2}\widehat{\mathbf{H}}\mathbf{D}\mathbf{R}_{\mathbf{X}_{\rm d}}\mathbf{D}\widehat{\mathbf{H}}^{\dagger}\right| \right]. 
 \label{eq:R_begin}
\end{equation}

\subsection{Problem Statement}  
We are interested in the throughput limit of the considered training-based multiuser MIMO system for a given number of antennas at the BS and a given coherence time, i.e., both $N$ and $T$ are fixed. With \eqref{eq:R_begin}, the throughput maximization problem is formulated as follows:
 \begin{subequations}\label{eq:pro_formu}
\begin{align}
\underset{ \mathbf{X}_{\rm p},\{\gamma_k\}, \{\gamma_k^\prime\}, K,\alpha}{\mbox{maximize}}
& R \ {\rm in}\ (\ref{eq:R_begin})   \label{eq:pro_formu_a} \\ 
\mbox{subject to} 
~~~ &  \frac{1}{\alpha T}\lVert\mathbf{x}_{{\rm p},k}\rVert^2 = \gamma_k  P_0,~0\leq \alpha\leq 1 \label{eq:pro_formu_b} \\
 &  \alpha \gamma_k \!+\! (1 \!-\! \alpha)\gamma_k^\prime \leq\! 1, \gamma_k^\prime \geq 0, k\in \mathcal{I}_K.
\end{align}
\end{subequations}

A similar problem for a training-based conventional MIMO system has been previously studied in \cite{zheng2002communication,hassibi2003much, Coldrey07}. Particularly, it is known that the optimal $K$ is $K = \alpha T$ for conventional MIMO \cite{Coldrey07}. In this paper, we tackle the problem in a more challenging multiuser setup. A major difference is that, due to the near far effect, the large-scale fading coefficients $\{d_k\}$ of users are in general very different from each other. In this case, it is generally difficult to find the optimal training matrix $\mathbf{X}_{\rm p}$. Instead, we derive tight upper and lower bounds of the throughput. Furthermore, we derive the optimal system design for throughput maximization under the setting of uniform large-scale fading, i.e., $\mathbf{D} = d\mathbf{I}_K$. Interestingly, we will show that $K = \alpha T$ is not necessarily optimal for a training-based multiuser MIMO system, even in the setting of uniform large-scale fading.

\section{Throughput Upper Bound}\label{Section III}
In this section, we establish a useful throughput upper bound by relaxing the constraints of (\ref{eq:pro_formu}), as detailed below.

\subsection{Problem Relaxation}
To start with, we focus on the optimization of the pilot matirx by assuming that the other parameters $\{\gamma_k\}, \{\gamma_k^\prime\}, K$, and $\alpha$ are given. Then, problem  \eqref{eq:pro_formu} reduces to 
\begin{subequations} \label{eq:optprob_1}
\begin{align}
\underset{\mathbf{X}_{\rm p}}{\mbox{maximize\,\,\,\,}} &  R \ {\rm in}\ (\ref{eq:R_begin}) \label{eq:optprob_1_a}\\
\mbox{subject to\,\,\,\,} & (\mathbf{X}_{\rm p}\mathbf{X}_{\rm p}^\dagger)_\text{diag} = \mathbf{R_X}  \label{eq:optprob_1_b}
\end{align}
\end{subequations}
where
\begin{equation}
\mathbf{R}_{\mathbf{X}} = \text{diag}\{\alpha \gamma_1 P_0 T, \cdots, \alpha\gamma_K P_0 T\}.
\end{equation}
The expectation in (\ref{eq:R_begin}) is taken over $\mathbf{\widehat{H}}$. The randomness of $\mathbf{\widehat{H}}$ comes from  the randomness of $\mathbf{Y}_{\rm p}$.
We see from \eqref{eq:Y_p} that $\mathbf{Y}_{\rm p}$ is a zero-mean random matrix with covariance $\mathbf{X}_{\rm p}^\dagger \mathbf{D}^2\mathbf{X}_{\rm p}+N_0 \mathbf{I}_{\alpha T}$. Thus, $\mathbf{Y}_{\rm p}$ can be equivalently expressed as
\begin{equation}
\mathbf{Y}_{\rm p} = \mathbf{G}\left( \mathbf{X}_{\rm p}^\dagger\mathbf{D}^2\mathbf{X}_{\rm p}+N_0 \mathbf{I}_{\alpha T}\right)^{\frac{1}{2}},
\label{eq:Y_p_tilde1}
\end{equation}
where $\mathbf{G}\in \mathbb{C}^{N\times \alpha T}$ is a random matrix with the elements  independently drawn from $\mathcal{CN}(0,1)$. Combining \eqref{eq:estimateH} and  \eqref{eq:Y_p_tilde1}, we rewrite the sum rate in \eqref{eq:optprob_1_a}  as 
\begin{equation}\label{eq:R_X_tilde}
R = (1-\alpha)\mathbb{E}\left[\log\left|\mathbf{I}_N+\tfrac{1}{\sigma_v^2}\mathbf{G}\mathbf{\widetilde{X}}^\dagger\mathbf{D}\mathbf{R_{X_{\rm d}}D}\mathbf{\widetilde{X}}\mathbf{G}^\dagger \right| \right]
\end{equation} 
where the expectation is taken over $\mathbf{G}$, and 
\begin{eqnarray}
\mathbf{\widetilde{X}} \!\!&=&\!\! \mathbf{D}\mathbf{X}_{\rm p}(\mathbf{X}_{\rm p}^\dagger\mathbf{D}^2 \mathbf{X}_{\rm p}+N_0 \mathbf{I}_{\alpha T} )^{-\frac{1}{2}} \label{eq:X_tilde}\\
\sigma_v^2 \!\!&=&\!\! {\rm tr}\left\{\left(\mathbf{I}_K-\mathbf{\widetilde{X}}\mathbf{\widetilde{X}}^\dagger\right)\mathbf{D}^2\mathbf{R}_{\mathbf{X}_{\rm d}}\right\}+N_0. \label{sigma_v}
\end{eqnarray}
Note that (\ref{sigma_v}) is obtained by substituting (\ref{eq:MMSE_matrix}) into (\ref{eq:equivalent_noise_power}).

We now consider the following problem:
\begin{subequations} \label{eq:optprob_5}
\begin{align}
\underset{\mathbf{X}_{\rm p}}{\mbox{maximize\,\,\,\,}} &  R \ {\rm in}\ (\ref{eq:R_X_tilde})
\label{eq:optprob_5_a}\\
\mbox{subject to\,\,\,\,} & (\mathbf{\widetilde{X}}\mathbf{\widetilde{X}}^\dagger)_\text{diag} \leq \!\mathbf{D^2R_X}\!\left(N_0\mathbf{I}_K\!+\!\mathbf{D^2R_X}\right)^{-1}  \label{eq:optprob_5_b}
\end{align}
\end{subequations}
where the relation ``$\leq$" means {\it less than or equal to in an entry-by-entry manner}. The theorem below reveals that (\ref{eq:optprob_5}) is a relaxation of (\ref{eq:optprob_1}). The proof of Theorem \ref{thm1} is presented in Appendix \ref{A1}.
\begin{thm}\label{thm1}
The optimal sum rate of problem (\ref{eq:optprob_1}) is upper bounded by the optimal sum rate of (\ref{eq:optprob_5}).
\end{thm}

\subsection{Pilot Design}
We now present the solution to \eqref{eq:optprob_5} that serves as a throughput upper bound of the orginal problem (\ref{eq:optprob_1}). 

To proceed, we consider the following eigen-decomposition:
\begin{equation}
\mathbf{\widetilde{X}}^\dagger\mathbf{DR_{X_{\rm d}}D}\mathbf{\widetilde{X}} = \mathbf{U}\bm{\Lambda}\mathbf{U}^\dagger
\end{equation}
where $\mathbf{U} \in \mathbb{C}^{\alpha T \times \alpha T}$ is a unitary matrix, and $\bm{\Lambda} = \text{diag}\{\lambda_1,\cdots,\lambda_{\text{min}(K, \alpha T)}, 0, \cdots, 0\} \in \mathbb{C}^{\alpha T\times\alpha T}$ is a diagonal matrix with $\lambda_i$ being the $i$th eigenvalue of $\mathbf{\widetilde{X}}^\dagger\mathbf{DR'_XD}\mathbf{\widetilde{X}}$. Note that $\mathbf{G}$ is unitarily invariant since the elements of $\mathbf{G}$ are i.i.d. Gaussian. That is, $\mathbf{GU}$ has the same distribution as $\mathbf{G}$ does. Therefore, the throughput in (\ref{eq:R_X_tilde}) can be equivalently written as
\begin{equation}\label{eqxqq1}
R = (1-\alpha)\mathbb{E}\left[\log\left|\mathbf{I}_N+\tfrac{1}{\sigma_v^2} \mathbf{G}\bm{\Lambda}\mathbf{G}^\dagger \right| \right].
\end{equation}
Then, problem (\ref{eq:optprob_5}) can be equivalently written as
\begin{subequations} \label{eq:optprob_52}
\begin{align}
\underset{\mathbf{X}_{\rm p}}{\mbox{maximize\,\,\,\,}} &  R \ {\rm in}\ (\ref{eqxqq1})
\label{eq:optprob_52_a}\\
\mbox{subject to\,\,\,\,} & (\mathbf{R}_{\mathbf{X}_{\rm d}}^\frac{1}{2}\mathbf{D\widetilde{X}}\mathbf{\widetilde{X}}^\dagger\mathbf{DR}_{\mathbf{X}_{\rm d}}^\frac{1}{2})_\text{diag} \nonumber \\ 
&\ \ \ \leq \mathbf{R'_XD^4R_X}\left(N_0\mathbf{I}+\mathbf{D^2R_X}\right)^{-1}. \label{eq:optprob_52_b}
\end{align}
\end{subequations}
The optimal solution to problem (\ref{eq:optprob_52}) is presented below, with the proof given in Appendix \ref{A2}.

\begin{thm} \label{thm:Optimal_pilots_content} 
For $K\leq \alpha T$, the optimal $\mathbf{X}_{\rm p}$ to \eqref{eq:optprob_52} satisfies 
\begin{equation}\label{eq5}
\mathbf{X}_{\rm p}\mathbf{X}_{\rm p}^{\dagger} = \mathbf{R_X}.
\end{equation} 
For $K > \alpha T$, the optimal $\mathbf{X}_{\rm p}$ to \eqref{eq:optprob_52} satisfies the following conditions:
\begin{itemize}
\item[1)] $(\mathbf{\widetilde{X}}\mathbf{\widetilde{X}}^\dagger\mathbf{D}^2\mathbf{R_{X_{\rm d}}})_\text{diag} = \mathbf{R_{X_{\rm d}}D^4R_X}\left(N_0\mathbf{I}_K+\mathbf{D^2R_X}\right)^{-1}$;
\item[2)] $\bm{\lambda}(\mathbf{\widetilde{X}}\mathbf{\widetilde{X}}^\dagger\mathbf{D}^2\mathbf{R_{X_{\rm d}}})$, with the last $(K - \alpha T)$ entries being zeros, is the smallest vector that majorizes the diagonal of $\mathbf{R_{X_{\rm d}}D^4R_X}\left(N_0\mathbf{I}_K+\mathbf{D^2R_X}\right)^{-1}$.\footnote{How to determine this smallest vector is elaborated in Remark \ref{rmk2} and Appendix \ref{appendix A}.}
\end{itemize}  \label{eq:Xp}
\end{thm}

\begin{rmk}\label{rmk1}
We now describe an explicit approach to construct the optimal $\mathbf{X}_{\rm p}$ in Theorem \ref{thm:Optimal_pilots_content}. For the case of  $K\leq \alpha T$, the optimal $\mathbf{X}_{\rm p}$  can be represented as $\mathbf{X}_{\rm p} = \mathbf{R}_{\mathbf{X}}^{\frac{1}{2}}\mathbf{U}$ where $\mathbf{U} \in \mathbb{C}^{K\times \alpha T}$ is an arbitrary orthonormal matrix satisfying $\mathbf{UU}^\dagger = \mathbf{I}_K$. The construction of optimal $\mathbf{X}_{\rm p}$ for the case of  $K> \alpha T$ is more involving. First, we have the following equalities:
\begin{subequations}\label{eq2}
\begin{eqnarray}
(\mathbf{\widetilde{X}}\mathbf{\widetilde{X}}^\dagger\mathbf{D}^2\mathbf{R}_{\mathbf{X}_{\rm d}})_\text{diag} \!\!\!\!&=&\!\!\!\! (\mathbf{R}^\frac{1}{2}_{\mathbf{X}_{\rm d}}\mathbf{D}\mathbf{\widetilde{X}}\mathbf{\widetilde{X}}^\dagger\mathbf{D}\mathbf{R}^\frac{1}{2}_{\mathbf{X}_{\rm d}})_\text{diag}\\
\bm{\lambda}(\mathbf{\widetilde{X}}\mathbf{\widetilde{X}}^\dagger\mathbf{D}^2\mathbf{R_{X_{\rm d}}})\!\!\!\! &=&\!\!\!\! \bm{\lambda}(\mathbf{R}^\frac{1}{2}_{\mathbf{X}_{\rm d}}\mathbf{D}\mathbf{\widetilde{X}}\mathbf{\widetilde{X}}^\dagger\mathbf{D}\mathbf{R}^\frac{1}{2}_{\mathbf{X}_{\rm d}})
\end{eqnarray}
\end{subequations}
where the second equality follows by noting the fact that matrices $\mathbf{AB}$ and $\mathbf{BA}$ share the same set of nonzero eigenvalues. Then, from Theorem 4.3.32 in \cite{horn}, for any two vectors $\mathbf{x} \succ \mathbf{y}$, a Hermitian matrix with $\mathbf{x}$ being the eigenvalues and $\mathbf{y}$ being the diagonal can be explicitly constructed. This, together with (\ref{eq2}) and the two conditions in Theorem \ref{thm:Optimal_pilots_content}, ensures that the Hermitian matrix $\mathbf{R}^\frac{1}{2}_{\mathbf{X}_{\rm d}}\mathbf{D}\mathbf{\widetilde{X}}\mathbf{\widetilde{X}}^\dagger\mathbf{D}\mathbf{R}^\frac{1}{2}_{\mathbf{X}_{\rm d}}$ indeed exists and can be explicitly constructed. Thus, we obtain $\mathbf{\widetilde{X}}\mathbf{\widetilde{X}}^\dagger$. With the definition in (\ref{eq:X_tilde}), we can determine $\mathbf{DX}_{\rm p} = \mathbf{U}_{\rm p}\mathbf{\Sigma}_{\rm p} \mathbf{V}_{\rm p}^\dagger$ based on the eigen-decomposition of  $\mathbf{\widetilde{X}}\mathbf{\widetilde{X}}^\dagger$, and hence $\mathbf{X}_{\rm p}$ is constructed.
\end{rmk}

\begin{rmk}\label{rmk2}
Another issue with Theorem \ref{thm:Optimal_pilots_content} is to determine the \it{smallest} \rm $\bm{\lambda}(\mathbf{\widetilde{X}}\mathbf{\widetilde{X}}^\dagger\mathbf{D}^2\mathbf{R_{X_{\rm d}}})$, with the last $(K - \alpha T)$ entries being zeros, that majorizes the diagonal of $\mathbf{R_{X_{\rm d}}D^4R_X}\left(N_0\mathbf{I}_K+\mathbf{D^2R_X}\right)^{-1}$. Without loss of generality, denote this smallest vector by $\bm{\lambda}^*$. Then, the word ``smallest'' means that, for any vector $\bm{\lambda}$ with the last $(\alpha T - K)$ entries being zeros, if $\bm{\lambda}$ majorizes the diagonal of $\mathbf{R_{X_{\rm d}}D^4R_X}\left(N_0\mathbf{I}_K+\mathbf{D^2R_X}\right)^{-1}$, then $\bm{\lambda}$ majorizes $\bm{\lambda}^*$. The explicit construction of $\bm{\lambda}$ is presented in Appendix \ref{appendix A}.
\end{rmk}

The following is an immediate result of Theorems \ref{thm1} and \ref{thm:Optimal_pilots_content}.
\begin{cor}\label{corollary 1}
For $K\leq \alpha T$, the optimal $\mathbf{X}_{\rm p}$ to \eqref{eq:optprob_1} satisfies $\mathbf{X}_{\rm p}\mathbf{X}_{\rm p}^{\dagger} = \mathbf{R_X}$.
\end{cor}
\begin{IEEEproof}
	From Theorems \ref{thm1} and \ref{thm:Optimal_pilots_content}, we see that $\mathbf{X}_{\rm p}$ satisfying $\mathbf{X}_{\rm p}\mathbf{X}_{\rm p}^{\dagger} = \mathbf{R_X}$ provides a throughput upper bound for problem (\ref{eq:optprob_1}). Further, it can be readily verified that such an $\mathbf{X}_{\rm p}$ falls in the feasible region of (\ref{eq:optprob_1}). Therefore, $\mathbf{X}_{\rm p}$ satisfying $\mathbf{X}_{\rm p}\mathbf{X}_{\rm p}^{\dagger} = \mathbf{R_X}$ achieves the optimum of problem (\ref{eq:optprob_1}).
\end{IEEEproof}

\begin{rmk}\label{rmk3}
For the case of $K> \alpha T$, the optimal solution to problem (\ref{eq:optprob_5}) in Theorem \ref{thm:Optimal_pilots_content} in general only gives a throughput upper bound of problem (\ref{eq:optprob_1}). The reason is that the equality in (\ref{eq:optprob_5_b}) holds when achieving the optimal solution of (\ref{eq:optprob_5}), which generally goes beyond the feasible region of (\ref{eq:optprob_1}).
\end{rmk}

\subsection{Optimization of $\{\gamma_k\}$ and $\{\gamma_k^\prime\}$}
With the pilot design in Theorem \ref{thm:Optimal_pilots_content}, we proceed to the optimization of the power coefficients $\{\gamma_k\}$ and $\{\gamma_k^\prime\}$ for any given values of $K$ and $\alpha$. 

\subsubsection{The Case of $K\leq\alpha T$}
For $K\leq\alpha T$, the pilot design in (\ref{eq5}) is optimal. Then, the throughput is given by (\ref{eqxqq1}) with
\begin{eqnarray}
	\sigma_v^2 &=& \sum_{k = 1}^{K}\frac{\gamma_k^\prime d_k^2P_0}{1+\alpha\gamma_k\rho_0d_k^2T} + N_0 \label{eqq2}\\
\lambda_k &=& \frac{\alpha\gamma_k\gamma_k^\prime\rho_0P_0 d_k^4 T}{1+\alpha\gamma_k\rho_0 d_k^2 T}, \ {\rm for}\ k = 1,\cdots, K\label{eqi1}\\
\rho_0 &=& \frac{P_0}{N_0}.
\end{eqnarray}
With the above, the optimization problem can be written as
\begin{subequations} \label{eq:opt_30}
\begin{align}
\underset{\{\gamma_k\},\{\gamma^\prime_k\}}{\mbox{maximize\,\,\,\,}} &  R \ {\rm in} \ (\ref{eqxqq1})
\label{eq:opt_30_a}\\
\mbox{subject to\,\,\,\,} & \alpha\gamma_k+(1-\alpha)\gamma^\prime_k\leq 1, \label{eq:opt_30_b}\\
& \gamma_k \geq 0, \gamma^\prime_k \geq 0,\ k \in \mathcal{I}_K. \label{eq:opt_30_c}
\end{align}
\end{subequations}
We have the following two observations. First, for any given $\gamma_k^\prime \neq 0$, $R$ in (\ref{eqxqq1}) is monotonically increasing in $\gamma_k$. Thus, the optimal $\gamma_k$ satisfies $\alpha\gamma_k+(1-\alpha)\gamma^\prime_k = 1$ for any $\gamma_k^\prime \neq 0$. Second, for $\gamma_k^\prime = 0$, $R$ is invariant to $\gamma_k$. Therefore, problem (\ref{eq:opt_30}) reduces to
\begin{subequations} \label{eq:optprob_301}
\begin{align}
\underset{\{\gamma_k\}}{\mbox{maximize\,\,\,\,}} &  R \ {\rm in} \ (\ref{eqxqq1}) \label{eq:optprob_301_a}\\
\mbox{subject to\,\,\,\,} & 0\leq \gamma_k\leq \frac{1}{\alpha},\ k \in \mathcal{I}_K \label{eq:optprob_301_b}
\end{align}
\end{subequations}
with $\gamma^\prime_k = \frac{1-\alpha\gamma_k}{1-\alpha}$. To solve the above problem, we introduce an auxiliary variable $t$, and convert (\ref{eq:optprob_301}) to the following form:
\begin{subequations} \label{eq:optprob_3012}
\begin{align}
\underset{\{\gamma_k\}, t}{\mbox{maximize\,\,\,\,}} &  (1-\alpha)\mathbb{E}\left[\log\left|\mathbf{I}_N+\tfrac{1}{t} \mathbf{G}\bm{\Lambda}\mathbf{G}^\dagger \right| \right]
\label{eq:optprob_3012_a}\\
\mbox{subject to\,\,\,\,} & 0\leq \gamma_k\leq \frac{1}{\alpha}, ~\sigma_v^2\leq t,\ k \in \mathcal{I}_K \label{eq:optprob_3012_b}
\end{align}
\end{subequations}
where $\gamma^\prime_k = \frac{1-\alpha\gamma_k}{1-\alpha}$; $\sigma_v^2$ and $\lambda_k$ are respectively given by (\ref{eqq2}) and (\ref{eqi1}). By noting that $R$ in (\ref{eqxqq1}) is monotonically increasing in $\sigma_v^2$, we can readily show that problem (\ref{eq:optprob_3012}) yields the same solution as (\ref{eq:optprob_301}).

We now show that for any given value of $t > 0$, problem (\ref{eq:optprob_3012}) is a convex problem. To see this, we first note that the target function (\ref{eq:optprob_3012_a}) is concave and monotonically increasing in $\{\lambda_k\}$, and that each $\lambda_k$ is concave in $\gamma_k$. Then, from the convexity composition rule, (\ref{eq:optprob_3012_a}) is a concave function of $\{\gamma_k\}$. Further, it can be readily shown that $\sigma_v^2$ is convex in $\{\gamma_k\}$. Thus, problem (\ref{eq:optprob_3012}) is convex for any given value of $t$, and can be solved by convex programming together with an exhaustive search over $t > 0$.

To get more intuitions, we present an explicit solution to (\ref{eq:optprob_301}) at high SNR, with the proof given in Appendix \ref{A3}.
\begin{thm}\label{thm7}
As $\rho_0$ tends to infinity, the asymptotically optimal $\{\gamma_k\}$ and $\{\gamma_k^\prime\}$ for problem (\ref{eq:optprob_301}) satisfy the following conditions: $\gamma_k = \gamma = \frac{1}{\alpha\left(1+\sqrt{(1-\alpha)\frac{T}{K}}\right)}$ and $\gamma_k^\prime = \gamma^\prime = \frac{1-\alpha\gamma}{1-\alpha}$ for $k = 1, \cdots, K$.
\end{thm}
\begin{rmk}
From Theorem \ref{thm7}, we see that the optimal power allocation for $K\leq \alpha T$ at high SNR is to allocate an equal amount of power for channel estimation for every user, no matter how good or bad the channel of a user is. Later in Section \ref{Section IV}, we will show that this is not the case when $K > \alpha T$. In fact, a portion of weak users should be deactivated in transmission when $K > \alpha T$.
\end{rmk}

\subsubsection{The Case of $K > \alpha T$}
We now consider the case of $K > \alpha T$. We first show that $\sigma_v^2$ in (\ref{sigma_v}) is still given by (\ref{eqq2}). To see this, we have
\begin{eqnarray}
\sigma_v^2 \!\!\!\!&=&\!\!\!\! \sum_{k=1}^{K}\gamma_k^\prime d_k^2 P_0 - {\rm tr}\left\{\mathbf{\widetilde{X}}\mathbf{\widetilde{X}}^\dagger\mathbf{D}^2\mathbf{R}_{\mathbf{X}_{\rm d}}\right\} + N_0\nonumber\\
&=&\!\!\!\! \sum_{k=1}^{K}\gamma_k^\prime d_k^2 P_0 \!-\! {\rm tr}\left\{\mathbf{R}_{\mathbf{X}_{\rm d}}\mathbf{D}^4\mathbf{R}_{\mathbf{X}}\left(N_0\mathbf{I}\!+\!\mathbf{D}^2\mathbf{R}_{\mathbf{X}}\right)^{-1}\right\}\!+\! N_0 \nonumber\\
&=&\!\!\!\! \sum_{k = 1}^{K}\frac{\gamma_k^\prime d_k^2P_0}{1+\alpha\gamma_k\rho_0d_k^2T} + N_0\label{eqq22}
\end{eqnarray}
where the first step follows from (\ref{sigma_v}), and the second step from Condition 1 of Theorem \ref{thm:Optimal_pilots_content}. Correspondingly, the throughput is written as
\begin{eqnarray}\label{eqq3}
R = (1-\alpha)\mathbb{E}\left[\log\left|\mathbf{I}_N+\frac{1}{\sigma_v^2}\mathbf{G}{\rm diag}\{\lambda_1,\cdots, \lambda_{\alpha T}\}\mathbf{G}^\dagger \right| \right]	
\end{eqnarray}
where $\{\lambda_k\}$ are determined by the fact that ${\bm \lambda} = (\lambda_1,\cdots,\lambda_{\alpha T}, 0,\cdots, 0)$ (with the last $K - \alpha T$ entries being zeros) is the minimum vector that majorizes the diagonal of $\mathbf{R_{X_{\rm d}}D^4R_X}\left(N_0\mathbf{I}+\mathbf{D^2R_X}\right)^{-1}$. For any given $\gamma_k^\prime$, $R$ is non-decreasing in $\gamma_k$. Therefore, the equality $\alpha\gamma_k+(1-\alpha)\gamma^\prime_k = 1$ holds at the maximizer.


The corresponding optimization problem is written as
\begin{subequations} \label{eqq4}
\begin{align}
\underset{\{\gamma_k\}, t}{\mbox{maximize\,\,\,\,}} &  \mathbb{E}\left[\log\left|\mathbf{I}_N+\frac{1}{t}\mathbf{G}{\rm diag}\{\lambda_1,\cdots, \lambda_{\alpha T}\}\mathbf{G}^\dagger \right| \right] \label{eqq4_a}\\
\mbox{subject to\,\,\,\,} & {\bm \lambda} \succ {\rm vdiag}\left\{\mathbf{R_{X_{\rm d}}D^4R_X}\left(N_0\mathbf{I}+\mathbf{D^2R_X}\right)^{-1}\right\} \label{eqq4_b}\\
& 0 \leq \gamma_k \leq \frac{1}{\alpha}, \gamma^\prime = \frac{1-\alpha\gamma_k}{1-\alpha}, \sigma_v^2 \leq t,\ k \in \mathcal{I}_K. \label{eqq4_c}
\end{align}
\end{subequations}
As the target function in (\ref{eqq4_a}) is Schur-concave in $\{\lambda_k\}$, the optimal ${\bm \lambda}$ is the minimum vector satisfying (\ref{eqq4_b}). From Appendix \ref{appendix A}, the optimal ${\bm \lambda}$ and ${\rm vdiag}\left\{\mathbf{R_{X_{\rm d}}D^4R_X}\left(N_0\mathbf{I}+\mathbf{D^2R_X}\right)^{-1}\right\}$ are linearly related by (\ref{eqq10}). Then, it can be shown that the target function in (\ref{eqq4_a}) is concave in $\{\gamma_k\}$, and that $\sigma_v^2$ is convex in $\{\gamma_k\}$. Therefore, for any given value of the auxiliary variable $t$, (\ref{eqq4}) is solvable using convex programming. Finally, the optimal solution to (\ref{eqq4}) can be found by an exhaustive search over $t$.

\subsection{Summary}
To summarize, the throughput upper bound developed in this section can be obtained as follows. For any given values of $\alpha$ and $K$, the optimal pilot matrix is given by Theorem \ref{thm:Optimal_pilots_content}. Then, for the case of $K \leq \alpha T$, the power allocation coefficients $\{\gamma_k\}$ and $\{\gamma_k^\prime\}$ can be determined by solving (\ref{eq:optprob_3012}) using convex programming plus a one-dimensional search; for the case of $K > \alpha T$, the power allocation coefficients $\{\gamma_k\}$ and $\{\gamma_k^\prime\}$ can be determined by solving (\ref{eqq4}). Finally, the optimal $\alpha$ and $K$ can be found by a two-dimensional exhaustive search.

%

\section{Throughput Lower Bound \label{Section IV}}
In this section, we establish a throughput lower bound. We start with the pilot design.
\subsection{Pilot Design}
The pilot design used in the throughput lower bound is presented in Algorithm \ref{alg2}. For the case of $K\leq \alpha T$, the pilot in Algorithm \ref{alg2} takes the optimal form given in (\ref{eq5}), which is the same as the case in the upper bound. The difference occurs in the case of $K > \alpha T$ where there is not enough time slots for the users to conduct orthogonal channel estimation. Intuitively, in this case, the channels of some distant users are so weak that allocating any time or power resource to these users leads to a degradation of the overall system performance. As such, a good strategy is to keep these weak users silent in transmission. Consequently, in Algorithm \ref{alg2}, the last $K-\alpha T$ diagonal elements of $\mathbf{R_X}$ are set to zeros. As the diagonal elements of $\mathbf{D}$ are arranged in a descending order, this implies that Algorithm \ref{alg2} selects $\alpha T$ active users with the largest large-scale fading coefficients $\{d_k\}$ in transmission. Later, we will show that this choice is asymptotically optimal at high SNR.

\begin{algorithm}
  \caption{Design of $\mathbf{X}_{\rm p}$ (Throughput Lower Bound)}
  \label{alg2}
  \textbf{Input:} $K, \alpha, \gamma_1, \cdots, \gamma_K.$\\
  \textbf{Output:} $\mathbf{X}_{\rm p}$.
  \begin{algorithmic}
  \IF{$K \leq \alpha T$}
  \STATE {Construct $\mathbf{X}_{\rm p}$ satisfying $\mathbf{X}_{\rm p}\mathbf{X}_{\rm p}^{\dagger}=\mathbf{R_X}$.}
  \ELSE[$K > \alpha T$] 
  \STATE {Construct $\mathbf{X}_{\rm p}$ satisfying $\mathbf{X}_{\rm p}\mathbf{X}_{\rm p}^{\dagger}=\mathbf{\overline{R}_X}$, where $\mathbf{\overline{R}_X}$ is obtained from $\mathbf{R_X}$ by setting its last $K-\alpha T$ diagonal elements to zeros.}
  \ENDIF
  \end{algorithmic}
\end{algorithm}


\subsection{Optimization of $\{\gamma_k\}$ and $\{\gamma_k^\prime\}$}
With the pilot design in Algorithm \ref{alg2}, we maximize the throughput over $\{\gamma_k\}$ and $\{\gamma_k^\prime\}$ as follows. For the case of $K\leq\alpha T$, $\mathbf{X}_{\rm p}$ in Algorithm \ref{alg2} is exactly the same as the one used in the upper bound. Therefore, the optimization problem of $\{\gamma_k\}$ and $\{\gamma_k^\prime\}$ is still given by (\ref{eq:optprob_3012}), with the high-SNR optimal solution given by Theorem \ref{thm7}. For the case of $K > \alpha T$, Algorithm \ref{alg2} chooses $\alpha T$ active users in transmission. Thus, the optimization problem is still given in the form of (\ref{eq:optprob_3012}). The only difference is that the optimization is now limited to the $K = \alpha T$ active users with the largest $d_k$ values, with the power coefficients corresponding to the inactive users set to $\gamma_k = \gamma_k^\prime = 0$. In the next subsection, we show that the above lower bound is asymptotically tight in the high SNR regime.

\subsection{Asymptotic Analysis}
In this subsection, we analyze the asymptotic behaviour of the above lower bound at high SNR. The main result is presented below, with the proof given in Appendix \ref{A4}.
\begin{thm}\label{thm6}
The pilot matrix $\mathbf{X}_{\rm p}$ given by Algorithm \ref{alg2} is asymptotically optimal in the sense of maximizing the throughput $R$ in (\ref{eq:R_X_tilde}), as $\rho_0$ goes to infinity. The optimal user selection for throughput maximization at high SNR is to select $\min\{K, \alpha T\}$ active users with the largest $d_k$ values in transmission. The corresponding optimal $\{\gamma_k\}$ and $\{\gamma_k^\prime\}$ are given by $\gamma_k = \gamma = \frac{1}{\alpha\left(1+\sqrt{\frac{(1-\alpha)T}{\min\{K,\alpha T\}}}\right)}$ and $\gamma_k^\prime = \frac{1-\alpha\gamma}{1-\alpha}$ if user $k$ is active; otherwise, $\gamma_k = \gamma_k^\prime = 0$.
\end{thm}

\begin{rmk}
The optimal power allocation derived in Theorem \ref{thm6} is similar to that in Theorem \ref{thm7}. In fact, Theorem \ref{thm6} extends the result of Theorem \ref{thm7} to include the case of $K > \alpha T$. Specifically, for $K > \alpha T$, the optimal design is to deactivate the $K-\alpha T$ users with the smallest large-scale fading coefficients in transmission.
\end{rmk}


\begin{figure}[!h]
\begin{centering}
\includegraphics[scale=0.63]{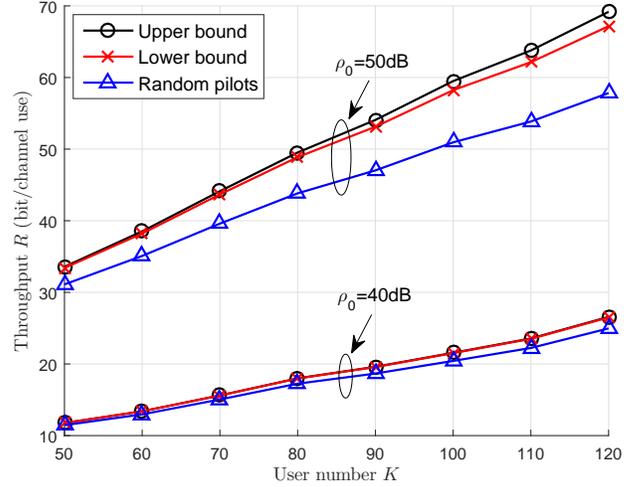}
\par\end{centering}
\centering{}\caption{The throughput against the number of users $K$ with SNR $\rho_0 = 40$ and $50$ dB.}\label{fig:throughput-vs-K}
\end{figure}

\begin{figure}[!h]
\begin{centering}
\includegraphics[scale=0.63]{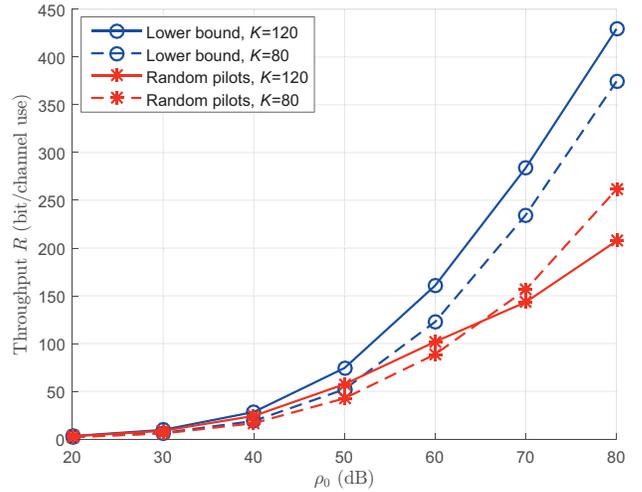}
\par\end{centering}
\centering{}\caption{The throughput against $\rho_0$ under various values of user number $K$.}\label{fig:throughput-vs-SNR-K}
\end{figure}

\begin{figure}[!h]
\begin{centering}
\includegraphics[scale=0.63]{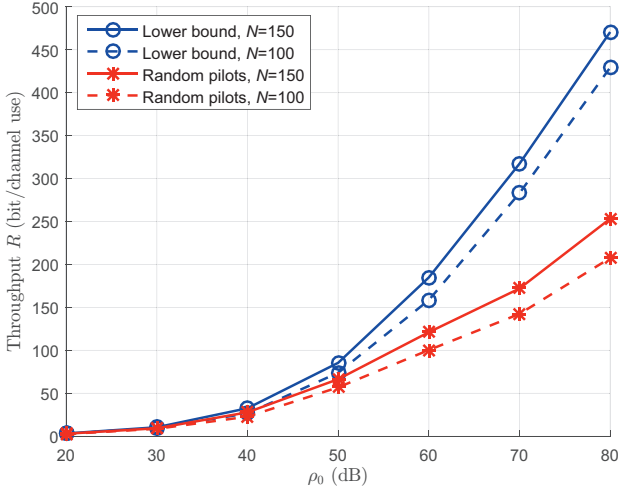}
\par\end{centering}
\centering{}\caption{The throughput against the SNR $\rho_0$ with various values of $N$.}\label{fig:throughput-vs-SNR-N}
\end{figure}

\begin{figure}[!h]
\begin{centering}
\includegraphics[scale=0.63]{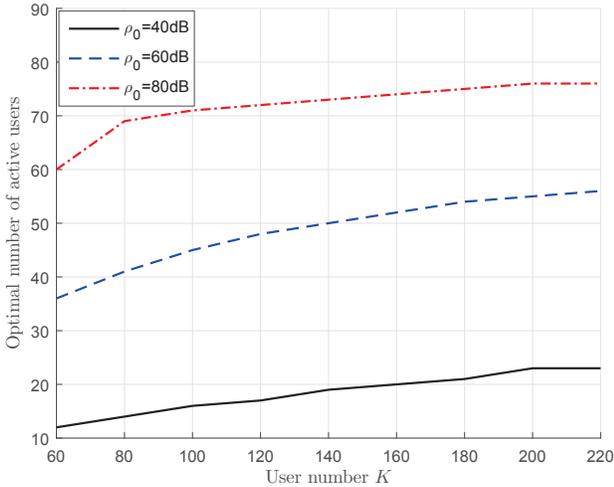}
\par\end{centering}
\centering{}\caption{The optimal number of active users against the total user number $K$ under various values of SNR $\rho_0$.}\label{fig:K-vs-optimalK}
\end{figure}

\subsection{Numerical Results}
We now present numerical results to examine the tightness of the established upper and lower bounds. \figref{fig:throughput-vs-K} demonstrates the throughput upper and lower bounds versus the user number $K$ with $\rho_0 = 40$ dB and $\rho_0 = 50$ dB. The upper bound is calculated based on the result in Section \ref{Section III}. The lower bound is given by Algorithm \ref{alg2}. The simulation settings are as follows: $N = 100$, $T = 200$. The large-scale fading coefficient $d_k$ of each user $k$ is modelled as $d_k = r_k^{-2}$ (corresponding to a large-scale fading exponent of $4$), where $r_k$ represents the distance between the base station and user $k$. It is assumed that users are uniformly distributed in a circle with radius 100 meters. Also, in simulation, the optimal $\alpha$ is obtained by exhaustive search for each given value of $K$. From \figref{fig:throughput-vs-K}, we see that the upper and lower bounds are very tight under various values of $\rho_0$ and $K$.

\figref{fig:throughput-vs-K} also includes the throughput behaviour for random pilots. For random pilots, every element of the pilot matrix $\mathbf{X}_{\rm p}$ is randomly drawn from a complex normal distribution $\mathcal{CN}(0, 1)$; then each row $k$ of $\mathbf{X}_{\rm p}$ is scaled to meet the power constraint of user $k$. The power allocation coefficients are set to $\gamma_1 = \cdots = \gamma_K = \gamma^\prime_1 = \cdots = \gamma^\prime_K = 1$. We see that the proposed optimal pilot design significantly outperforms the random pilot design, especially in the case of $\rho_0 = 50$ dB.

\figref{fig:throughput-vs-SNR-K} illustrates the throughput lower bound and the throughput for random pilots against the SNR $\rho_0$ with $K = 60$ and $K = 80$. The simulation settings follow those of \figref{fig:throughput-vs-K}. We see that the proposed lower bound significantly outperforms the random pilot design in the medium to high SNR regime. We also see that at high SNR, the lower bound performs better when the number of users increases from 60 to 80, while the opposite is observed for the random pilot design. The reason is as follows. For the lower bound, increasing the number of users provides more freedom to select the set of active users with better channels; however, for random pilot design, as all the users are active, more users imply higher interference.

\figref{fig:throughput-vs-SNR-N} is similar to \figref{fig:throughput-vs-SNR-K} but with different settings. Specifically, we set $K = 120$ and $T = 200$ and $N = 100, 150$. Again, we see that the proposed lower bound significantly outperforms the random pilot design. Also, we see that the throughput grows with the increase of $N$, due to the beamforming gain of the BS antenna array.

\figref{fig:K-vs-optimalK} illustrates the optimal number of active users against the number of users $K$ with SNR $\rho_0 = 40, 60, 80$ dB. The other settings follow \figref{fig:throughput-vs-K}. In \figref{fig:K-vs-optimalK}, the optimal number of active users may be significantly less than the number of available users $K$. We see that the optimal number of active users increase monotonically with $K$. Intuitively, the reason is that, with a larger $K$, there are more users with relatively good channels to be activated for transmission. Moreover, we also see from \figref{fig:K-vs-optimalK} that the optimal number of active users increase with SNR $\rho_0$.

\section{Throughput Optimization with Uniform Large-Scale Fading \label{Section V}}
In this section, we consider the throughput maximization when all the users are co-located, i.e., $\mathbf{D} = d\mathbf{I}_K$. We derive the optimal training design for this setup, and compare the optimal design with the upper and lower bounds in the preceding sections. Due to user symmetry, we always assume a common power allocation factor $\gamma$, i.e., $\gamma = \gamma_1 = \cdots = \gamma_K$.

\subsection{Optimal Pilot Design}
The following theorem gives the optimal pilot design under the assumption of uniform large-scale fading, with the proof presented in Appendix \ref{A5}.

\begin{thm} \label{thm3} Assume $d_1 = d_2 = \cdots = d_K = d$ and $\gamma_1 = \gamma_2 = \cdots = \gamma_K = \gamma$. Then, for $K\leq\alpha T$, the optimal training matrix $\mathbf{X}_{\rm p}$ to \eqref{eq:optprob_1} satisfies
\begin{subequations}
\begin{equation}
\mathbf{X}_{\rm p}\mathbf{X}_{\rm p}^{\dagger}=\alpha  \gamma P_0T \mathbf{I}_{K}; \label{eq:XpX_a} 
\end{equation}
For $K>\alpha T$, the optimal training matrix satisfies
\begin{align}
&\mathbf{X}_{\rm p}^{\dagger}\mathbf{X}_{\rm p}=\gamma P_0K \mathbf{I}_{\alpha T} \label{eq:XpX_b} \\
&  \lVert \mathbf{x}_{{\rm p},k}\rVert^2 = \alpha \gamma P_0T,  ~~~~~~~k\in \mathcal{I}_K.  \label{eq:XpX_c}
\end{align}\label{eq:XpX}
\end{subequations}
\end{thm}


\begin{rmk}
An explicit construnction of the optimal pilot matrix $\mathbf{X}_{\rm p}$ in Theorem \ref{thm3} is described as follows.  We focus on the case of $K> \alpha T$, as the case of $K\leq \alpha T$ is straightforward. For $K\!>\!\alpha T$, $\mathbf{X}_{\rm p}\in \mathbb{C}^{K\times\alpha T}$ is a tall matrix. To meet the conditions in \eqref{eq:XpX_b} and \eqref{eq:XpX_c} simultaneously, $\mathbf{X}_{\rm p}$ can be formed by extracting $\alpha T$ columns of the $K$-by-$K$ normalized discrete-Fourier-transform (or Hadamard) matrix.
\end{rmk}

\begin{rmk}
An implication of Theorem \ref{thm3} is that, if the optimal $K$ occurs at $K > \alpha T$, then the optimal training sequences (i.e., the rows of $\mathbf{X}_{\rm p}$) are not orthogonal to each other. Later, we will see that the optimal $K$ may occur at $K > \alpha T$ for the multiuser MIMO system in consideration. This is in contrast with the case of conventional MIMO where the optimal training sequences are always orthogonal.
\end{rmk}

\begin{rmk}
It is also interesting to compare the optimal pilot design in Theorem \ref{thm3} with the upper bound in Theorem \ref{thm:Optimal_pilots_content}. For the case of $K\leq \alpha T$, it can be readily shown that (\ref{eq5}) reduces to (\ref{eq:XpX_a}) by letting $\gamma_1 = \gamma_2 = \cdots = \gamma_K = \gamma$. Thus, both theorems give the same pilot design for $K\leq \alpha T$. What is more interesting is the case of $K> \alpha T$. In this case, it can be shown that the pilot design satisfying (\ref{eq:XpX_b}) and (\ref{eq:XpX_c}) in general does not meet the two conditions specified in Theorem \ref{thm:Optimal_pilots_content}. Therefore, Theorem \ref{thm:Optimal_pilots_content} only provides an upper bound even in the case of uniform large-scale fading, i.e., $d_1 = d_2 = \cdots = d_K = d$.
\end{rmk}

Based on Theorem \ref{thm3}, we simplify the throughput expression as follows. First note that $\sigma_v^2$ in (\ref{sigma_v}) can be rewritten as
\begin{subequations}
\begin{empheq}[left={\sigma_v^2 \!=\! \empheqlbrace}]{alignat=2}
&\frac{\gamma^{\prime} d^2P_0N_0K}{\alpha \gamma d^2P_{0}T+N_0}+N_0, &K\!\leq\! \alpha T\label{sigmav} \\
&\frac{\gamma d^2P_0 (K\!\!-\!\!\alpha T)\!\!+\!\!N_0}{\gamma d^2P_0K +N_0}\gamma^{\prime} d^2P_0K\!\!+\!\!N_0, &\ K\!>\!\alpha T
\end{empheq}
\end{subequations}
where 
\begin{equation}
\gamma^\prime = \frac{1-\alpha\gamma}{1-\alpha}.
\end{equation} 
It is worth noting that, for the upper bound in Theorem \ref{thm:Optimal_pilots_content}, the expression of $\sigma_v^2$ is given by (\ref{sigmav}) for both $K\leq \alpha T$ and $K > \alpha T$.

With the above, the throughput in (\ref{eq:R_X_tilde}) can be expressed as
\begin{equation}
R = \left(1{-}\alpha\right) { \mathbb{E}}\left[  \log\left|\mathbf{I}_{N}\!+\tau \widetilde{\mathbf{G}}\widetilde{\mathbf{G}}^{\dagger}\right| \right]\label{eq11}
\end{equation}
where the elements of $\mathbf{\widetilde{G}}\in\mathbb{C}^{N\times\min\{K, \alpha T\}}$ are independently drawn from $\mathcal{CN}(0,1)$, and
\begin{small}
\begin{subequations}\label{eq31}
\begin{empheq}[left={\tau \!=\! \empheqlbrace}]{alignat=2}
&\frac{\alpha\gamma\gamma^{\prime}\rho_0^2 d^4T}{\rho_0d^2(\gamma^\prime K+\alpha\gamma T)+1}, &K\!\leq\! \alpha T \\
&\frac{\gamma\gamma^\prime\rho_0^2 d^4K}{\gamma\gamma^\prime\rho_0^2 d^4K(K\!\!-\!\!\alpha T)\! +\!\rho_0d^2K(\gamma\!+\!\gamma^\prime)\!+\!1}, &\ K\!>\!\alpha T
\end{empheq}
\end{subequations}
\end{small}with SNR $\rho_0 = \frac{P_0}{N_0}$. Then, the optimization problem can be rewritten as
\begin{subequations}\label{eq:prob_without_Xp}
\begin{alignat}{3}
\underset{\gamma,K,\alpha}{\mbox{maximize}} & ~~~~R \ {\rm in} \ (\ref{eq11}) \label{eq:prob_without_Xp_a} \\
\mbox{subject to}&~~~~0\leq \alpha \leq 1, ~0\leq \gamma \leq \tfrac{1}{\alpha}.  \label{eq:prob_without_Xp_b} 
\end{alignat}
\end{subequations}

\subsection{Optimization over $\gamma$}
We first optimize the power coefficient $\gamma$. For any fixed values of $K$ and $\alpha$, the coefficient $\gamma$ is only related to $\tau$ in \eqref{eq11}. Thus, the optimization problem with respect to $\gamma$ can be written as 
\begin{subequations}\label{eq:power_all_pro}
\begin{align}
\underset{\gamma}{\mbox{maximize}}&~~~~ \tau \ {\rm in}\ (\ref{eq31}) \\
\mbox{subject to} &~~~~ 0\leq \gamma\leq {1}/{\alpha}
\end{align} 
where $\tau$ is defined below (\ref{eq11}). The solution to \eqref{eq:power_all_pro} is presented below.
\end{subequations}

\begin{thm} \label{thm:optimal_power}
The optimal  $\gamma$ to \eqref{eq:power_all_pro}  is given by
\begin{subequations}  
	\begin{empheq}[left={\gamma^{\rm opt}=\empheqlbrace}]{alignat=2}
	&\frac{1}{\alpha (1+\sqrt{1-\mu_1})}, ~~K\leq \alpha T \label{eq:power_all_prt1_a} \\ 
	&\frac{1}{\alpha (1+\sqrt{1-\mu_2})},  ~~K>\alpha T,  \label{eq:power_all_prt1_b}
	\end{empheq} \label{eq:power_all_prt1}
\end{subequations}
where $\mu_1 = \tfrac{\rho_0d^2(K-(1-\alpha )T )}{1-\alpha+\rho_0d^2K}$ and  $\mu_2 = \tfrac{\rho_0d^2(2\alpha -1)K}{\alpha (1-\alpha+\rho_0d^2K)}$.
\end{thm}
\begin{IEEEproof}
The optimal $\gamma$ for \eqref{eq:power_all_pro} is readily obtained by solving the Karush-Kuhn-Tucker (KKT) conditions \cite{Boyd2004}. 
\end{IEEEproof}
\begin{rmk}
We note that the optimal $\gamma$  given by  \eqref{eq:power_all_prt1} is continuous at $K=\alpha T$.
\end{rmk}

\subsection{Optimization over $K$ \label{sub:Optimizing-over-K}}
We now consider the optimization of $K$ for given $\alpha$. We have the following main result, with the proof given in Appendix \ref{A6}.

\begin{thm}{\label{lem:KleqalphaT}}
For any given $\alpha\in [0, 1]$, the optimal $K = K^{\rm opt}$ for \eqref{eq:prob_without_Xp} is given by
\begin{equation}\label{Kopt}
K^{\rm opt} = \max\left\{\frac{x^*}{\rho_0d^2}, \alpha T\right\},
\end{equation} 
where $x^*$ is the root of
\begin{small}
\begin{eqnarray}
f(x) = -x^2-x\sqrt{\alpha-\alpha^2}\left(\sqrt{\frac{x+\alpha}{x+1-\alpha}}+\sqrt{\frac{x+1-\alpha}{x+\alpha}}\right)\nonumber\\
+2\sqrt{(\alpha-\alpha^2)(x+\alpha)(x+1-\alpha)}+2(\alpha-\alpha^2).\label{eq12}
\end{eqnarray}
\end{small} 
\end{thm}

For a general SNR $\rho_0$, we have no explicit expression of the optimal $K$ in terms of $\alpha$. But we have a closed-form expression of the optimal $K$ in the high SNR regime, as presented below.

\begin{cor} \label{corollary 2}
For any given $\alpha \in [0, 1]$, the optimal $K$ to \eqref{eq:prob_without_Xp} satisfies 
\begin{equation}
K \rightarrow \alpha T,~~~~ \mbox{as }\rho_0 \rightarrow \infty.
\label{eq:opt_K}
\end{equation}
\end{cor}
\begin{IEEEproof}
The corollary holds by letting $\rho_0 \rightarrow \infty$ in (\ref{Kopt}).
\end{IEEEproof}

\begin{rmk} \label{rmk:4}
The tradeoff involved in optimizing $K$ are elaborated as follows.
On one hand, for the data transmission phase, it is known from the information theory that the throughput of the considered multiuser MIMO channel increases unboundedly as the user number $K$ tends to infinity, provided that the user channels are perfectly known.
On the other hand, for the training phase, the channel estimation accuracy decreases in a growing $K$, as more channel coefficients need to be estimated as $K$ increases. From Theorem \ref{lem:KleqalphaT}, the optimal tradeoff occurs when $K\geq \alpha T$. Further, Corollary \ref{corollary 2} states that the optimal $K$ tends to $\alpha T$ in the high SNR regime.
\end{rmk}

\begin{rmk}\label{rmk:5}
The remaining issue is to optimize $\alpha$. Though it is difficult to derive an explicit expression, the optimal $\alpha$ can be readily obtained by using an exhaustive search over $[0, 1]$.
\end{rmk}

%

\subsection{Large System Analysis \label{sec:Asymototic-analysis}}
In this subsection, we present an approximate expression of the throughput by using the random matrix theory. Let $\lambda$ be a non-zero eigenvalue of $\frac{1}{N}\widetilde{\mathbf{G}}\widetilde{\mathbf{G}}^\dagger$. Then
\begin{propo}{\label{propo:eig_dis}}
As $N, K, T\rightarrow \infty $ with fixed ratios of $K/N \!=\! \beta$ and $K/T = \omega$, the asymptotic distribution of $\lambda$ is given by
\begin{subequations}\label{eq:f_beta}
\begin{empheq}[left={f_{\beta,\omega}\left(\lambda\right){=}\empheqlbrace}]{alignat=2}
&\tfrac{\sqrt{\left(\lambda{-}a\right)^{+}\left(b-\lambda\right)^{+}}}{2\pi\lambda},   ~&~~~ \omega \leq \alpha  \label{eq:f_beta_a} \\
&\tfrac{\sqrt{\left(\lambda-a'\right)^{+}\left(b^\prime-\lambda\right)^{+}}}{2\pi\lambda},  ~& ~~~\omega>\alpha \label{eq:f_beta_b},
\end{empheq}
\end{subequations}
where  $a=(1{-}\sqrt{\beta})^{2}$, $b=(1{+}\sqrt{\beta})^{2}$, $a'=(1-\sqrt{\alpha\beta/\omega})^{2}$ and $b'=(1{+}\sqrt{\alpha\beta/\omega})^{2}$.
\end{propo}
\begin{IEEEproof}
Recall that the elements of $\mathbf{\widetilde{G}}\in\mathbb{C}^{N\times\min\{K, \alpha T\}}$ are independently drawn from $\mathcal{CN}(0,1)$. Then, the proof is immediate by noting Theorem 2.35 in \cite{tulino2004random}.
\end{IEEEproof}

With Proposition \ref{propo:eig_dis}, the throughput in \eqref{eq11} can be approximated by \eqref{eq:R_asyform}. It is known that the large-system approximation is accurate even when the system parameters $N, K, T$ are relatively small \cite{tulino2004random}. Compared with \eqref{eq11}, an advantage of \eqref{eq:R_asyform} is that no Monte Carlo simulation is required in evaluating the throughput. Also, \eqref{eq:R_asyform} provides a simpler analytical throughput characterization when applying the analytical methods in this work to a massive MIMO setup.

\begin{figure*}[ht]
\begin{subequations} 
	\small
	\begin{empheq}[left={R\ {=}\empheqlbrace}]{alignat=2}
	&{\left(1{-}\alpha\right) N} \int_{a}^{b}\log \left(1{+}\frac{\alpha \gamma\gamma' \rho_0^{2}d^4 T N}{\rho_0 d^2\left(\gamma'K{+}\alpha\gamma T\right){+}1}\lambda \right) \frac{\sqrt{\left(\lambda{-}a\right)\left(b{-}\lambda\right)}}{2\pi\lambda}d\lambda, & ~~K\leq\alpha T  \label{eq:R_asyform_a} \\ 
	&{\left(1{-}\alpha\right) N}  \int_{a'}^{b'}\log \left(1{+}\frac{\gamma \gamma'\rho_0^2 d^4 KN}{\gamma'\gamma\rho_0^{2}d^4K\left(K{-}\alpha T\right){+}\left(\gamma'{+}\gamma\right)\rho_0d^2K{+}1}\lambda \right)\frac{\sqrt{\left(\lambda{-}a'\right)\left(b'{-}\lambda\right)}}{2\pi\lambda}d\lambda, & ~~~K>\alpha T.  \label{eq:R_asyform_b}
	\end{empheq} \label{eq:R_asyform}
\end{subequations} 
\hrulefill
\end{figure*}

\begin{figure}[tbh]
\begin{centering}
\includegraphics[scale=0.64]{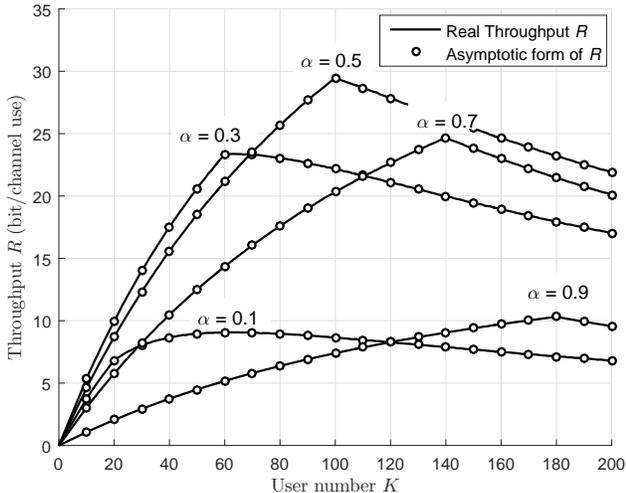}
\par\end{centering}
\caption{The comparison of the real throughput  and its asymptotic expression.}\label{fig:Approximation-error}
\end{figure}

\subsection{Numerical Results}
We now present numerical results to verify the analysis. \figref{fig:Approximation-error} illustrates the throughputs given by \eqref{eq:R_asyform} and \eqref{eq11} against $K$ with various values of $\alpha$. The settings are $\gamma =1$, $\beta = K/N = 1$, $\rho_0 = -18$ dB, $N = 100$, and $T=200$. In \figref{fig:Approximation-error}, we see that the two curves always coincide with each other. This demonstrates that \eqref{eq:R_asyform} is indeed a good approximation of the throughput in \eqref{eq11}. 

\begin{figure}[!h]
\begin{centering}
\includegraphics[scale=0.63]{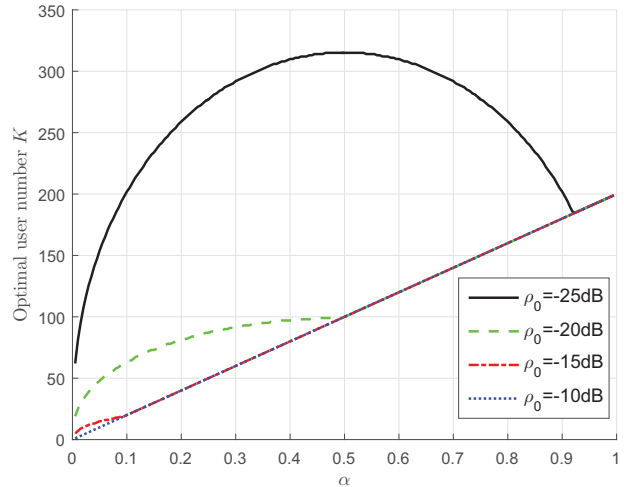}
\par\end{centering}
\centering{}\caption{The optimal number of users $K$ vs the time allocation factor $\alpha$ under various values of SNR $\rho_0$.}\label{fig:Optimal_K}
\end{figure}

\figref{fig:Optimal_K} illustrates the optimal user number $K$ versus $\alpha$ under various SNR values. The simulation settings are as follows: $N = 300$, $T = 200$, $d_1 = \cdots = d_K = d =1$. We see that the optimal $K$ is always no less than $\alpha T$, and it converges to $\alpha T$ for an arbitrary value of $\alpha$ when $\rho_0$ goes to infinity. This is in well agreement with Theorem \ref{thm3} and Corrolary \ref{corollary 2}.

\begin{figure}[!h]
\centering
\includegraphics[scale=0.66]{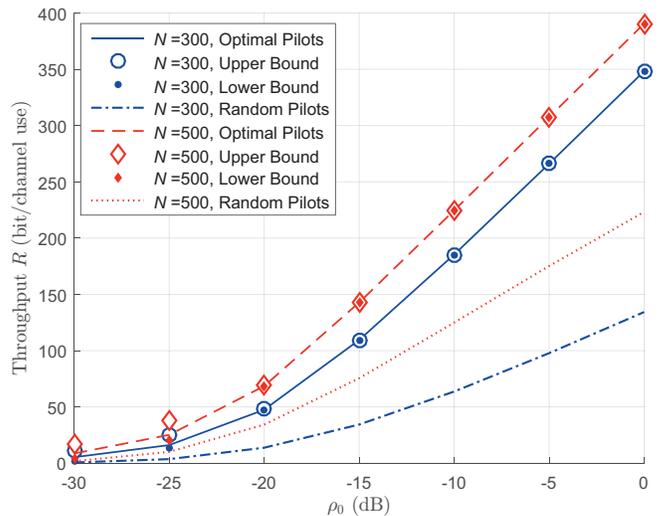}
\caption{The optimal throughput against $\rho_0$ under different values of $N$.}\label{fig:R_N_T1000}
\end{figure}

\figref{fig:R_N_T1000} illustrates the throughput against the SNR $\rho$ with $T = 200$ and $N = 300, 500$. The optimal pilot design and the random pilot design are considered in simulation. The optimal pilot design is given by Theorem \ref{thm3}. For random pilot design, we set $\gamma = \gamma^\prime = 1$ and $K = \alpha T$; each element of the pilot matrix $\mathbf{X_{\rm p}}$ is independently drawn from a complex Gaussian distribution, and then each row of $\mathbf{X_{\rm p}}$ is scaled to meet the power constraint for each user. In both designs, the optimal $\alpha$ is obtained through an exhaustive one-dimensional search. We see that the optimal pilot design significantly outperforms the random pilot design especially in the medium to high SNR regime. We see that the throughput increases with $N$. This is because a larger $N$ implies a higher power gain of the receiving antenna array. Moreover, we also include the upper and lower bounds developed in Sections \ref{Section III} and \ref{Section IV}. We see that these bounds are very tight in the medium to high SNR regime.

\section{Conclusions \label{Section VI}}
In this paper, we analyzed the tradeoff between the channel estimation quality and the information throughput of a training-based multiuser MIMO system. We first studied the pilot design, and established upper and lower bounds for throughput maximization. We then considered the optimization of power coefficients for training and data transmission. We showed that the established lower bound is asymptotically optimal in the high SNR regime.

Our analysis revealed that the optimal training design strategies for a multiuser MIMO system are very different from those for conventional MIMO. For example, we showed that the optimal training sequences for throughput maximization in a multiuser MIMO system are not necessarily orthogonal to each other. Also, due to the near-far effect, the optimal training design strategy for throughput maximization is to inactivate a portion of users with the weakest channels in transmission. These findings will provide insights and guidance for the practical design of a training-based multiuser system.

Future research may arise in a number of directions. For example, the work in this paper was focused on the case of a single BS with multiple antennas, without considering the interference from other base stations. BS cooperation can significantly mitigate such interference; see, e.g., \cite{Lozano2014, CRAN2, Hermi} and the references therein. How to characterize the tradeoff between the channel estimation overhead and the information throughput in a cellular network with BS cooperation will be a challenging research topic deserving future endeavour.

\appendices
\section{Proof of Theorem \ref{thm1}}\label{A1}
As problems (\ref{eq:optprob_1}) and (\ref{eq:optprob_5}) have the same objective function, it suffices to show that the feasible region of (\ref{eq:optprob_5}) contains that of (\ref{eq:optprob_1}), or in other words, to show that (\ref{eq:optprob_5_b}) is implied by (\ref{eq:optprob_1_b}). To proceed, let the singular value decomposition of $\mathbf{DX}_{\text{p}}$ be
\begin{equation}\label{eq:Xp_SVD} 
\mathbf{DX}_{\rm p} = \mathbf{U}_{\rm p}\mathbf{\Sigma}_{\rm p} \mathbf{V}_{\rm p}^\dagger
\end{equation}
where $\mathbf{\Sigma}_{\rm p}\in \mathbb{C}^{K\times \alpha T}$ is a diagonal matrix with non-negative diagonal elements, and  $\mathbf{U}_{\rm p}\in \mathbb{C}^{K\times K}$ and $\mathbf{V}_{\rm p}\in\mathbb{C}^{\alpha T\times\alpha T}$ are unitary matrices. Then, (\ref{eq:optprob_1_b}) can be equivalently expressed as
\begin{equation}
\sum_{j=1}^K \left|u_{k,j}\right|^2 \sigma_{{\rm p}, j}^2 = \alpha d_k^2\gamma_k P_0 T,~ k\in \mathcal{I}_K \label{eq1}
\end{equation}
where $u_{k,j}$ is the $(j,k)$th element of $\mathbf{U}_{\rm p}$; $\sigma_{{\rm p}, j}$ is the $j$th diagonal element of $\mathbf{\Sigma}_{\rm p}$ for $1\leq j \leq \text{min}(K, \alpha T)$ and $\sigma_{{\rm p}, j} = 0$ otherwise. Note that $\sum_{j=1}^K \left|u_{k,j}\right|^2 = 1$, and that $\frac{x}{N_0 + x}$ is a concave function of $x$. Then, we obtain
\begin{eqnarray}
\sum_{j=1}^K \left|u_{k,j}\right|^2 \frac{\sigma_j^2}{N_0+\sigma_j^2} \!\!&\leq &\!\! \frac{\sum_{j=1}^K \left|u_{k,j}\right|^2\sigma_j^2}{N_0+\sum_{j=1}^K \left|u_{k,j}\right|^2\sigma_j^2}\nonumber\\
\!\! &= &\!\! \frac{\alpha d_k^2\gamma_k P_0 T}{N_0 + \alpha d_k^2\gamma_k P_0 T},~ k\in \mathcal{I}_K\ \ \label{eqx3}
\end{eqnarray}
where the first step is from the Jensen's inequality, and the second step from (\ref{eq1}).
Noting that (\ref{eqx3}) is equivalent to (\ref{eq:optprob_5_b}), we conclude the proof of Theorem \ref{thm1}.

\section{Proof of Theorem \ref{thm:Optimal_pilots_content}}\label{A2}
For ease of disscussion, we first ignore the effect of $\sigma_v^2$ in throughput maximization. The minimization of $\sigma_v^2$ (so as to maximize the throughput) is discussed at the end of the proof. 

From Theorem 4.3.26 in \cite{horn}, we obtain
\begin{equation}
\bm{\lambda}(\mathbf{R}^\frac{1}{2}_{\mathbf{X}_{\rm d}}\mathbf{D\widetilde{X}}\mathbf{\widetilde{X}}^\dagger\mathbf{DR}^\frac{1}{2}_{\mathbf{X}_{\rm d}}) \succ \text{vdiag}\{\mathbf{R}_{\mathbf{X}_{\rm d}}^\frac{1}{2}\mathbf{D\widetilde{X}}\mathbf{\widetilde{X}}^\dagger\mathbf{DR}_{\mathbf{X}_{\rm d}}^\frac{1}{2}\}.\label{eq4}
\end{equation}
The definition of the majorization ``$\succ$'' can be found in Appendix \ref{appendix A}. From the matrix theory, $\mathbf{R}^\frac{1}{2}_{\mathbf{X}_{\rm d}}\mathbf{D\widetilde{X}}\mathbf{\widetilde{X}}^\dagger\mathbf{DR}^\frac{1}{2}_{\mathbf{X}_{\rm d}}$ and  $\mathbf{\widetilde{X}}^\dagger\mathbf{DR'_X}\mathbf{D\widetilde{X}}$ share the same set of nonzero eigenvalues. That is, $\bm{\lambda}(\mathbf{R}^\frac{1}{2}_{\mathbf{X}_{\rm d}}\mathbf{D\widetilde{X}}\mathbf{\widetilde{X}}^\dagger\mathbf{DR}^\frac{1}{2}_{\mathbf{X}_{\rm d}})$ and $\bm{\Lambda}$ have the same set of nonzero elements. Thus, problem (\ref{eq:optprob_5}) can be recast as to maximize the rate in (\ref{eqxqq1}) subject to (\ref{eq:optprob_52_b}) and (\ref{eq4}). As the rate in (\ref{eqxqq1}) is monotonically increasing in each $\lambda_i$, we see that the maximum of the problem occurs when (\ref{eq:optprob_52_b}) takes the equality. Therefore, problem (\ref{eq:optprob_5}) can be rewritten as
\begin{subequations}\label{eq:optprob_10} 
\begin{align}
\underset{\bm{\lambda}}{\mbox{maximize\,\,\,\,}} & R\ {\rm in}\ (\ref{eqxqq1}) \label{eq:optprob_10_a}\\
\mbox{subject to\,\,\,\,} & \bm{\lambda} \succ \text{vdiag}\{\mathbf{R_{X_{\rm d}}D^4R_X}\left(N_0\mathbf{I}_K+\mathbf{D^2R_X}\right)^{-1}\}. \label{eq:optprob_10_b}
\end{align}
\end{subequations} 
where $\bm{\lambda} = (\lambda_1, \cdots, \lambda_{K})$ for $K \leq \alpha T$, and $\bm{\lambda} = (\lambda_1, \cdots, \lambda_{\alpha T}, 0, \cdots, 0)$ for $K> \alpha T$.

We now solve problem (\ref{eq:optprob_10}) for two different cases, namely, $K\leq \alpha T$ and $K> \alpha T$. For the case of $K\leq \alpha T$, we have $\bm{\lambda} = (\lambda_1, \cdots, \lambda_{K})$. Note that $R$ in (\ref{eqxqq1}) is symmetric and concave with respect to $\lambda_1, \cdots, \lambda_{K}$. From Proposition C.2 in \cite{Majortheory}, $R$ in (\ref{eqxqq1}) is Schur concave. Thus, the optimal $\bm{\lambda}$ to problem (\ref{eq:optprob_10}) is given by $\bm{\lambda} = \text{vdiag}\{\mathbf{R_{X_{\rm d}}D^4R_X}\left(N_0\mathbf{I}_K+\mathbf{D^2R_X}\right)^{-1}\}$. This implies that the optimal $\mathbf{X}_p$ to (\ref{eq:optprob_5}) satisfies (\ref{eq5}), which proves the first half of the theorem.

For the case of $K> \alpha T$, as the rank of $\mathbf{R}^\frac{1}{2}_{\mathbf{X}_{\rm d}}\mathbf{D\widetilde{X}}\mathbf{\widetilde{X}}^\dagger\mathbf{DR}^\frac{1}{2}_{\mathbf{X}_{\rm d}}$ is at most $\alpha T$, we have $\bm{\lambda} = (\lambda_1, \cdots, \lambda_{\alpha T}, 0, \cdots, 0)\in \mathbb{C}^K$. Thus, unlike the case of $K\leq \alpha T$, we cannot set $\bm{\lambda}$ to $\bm{\lambda}  = \text{vdiag}\{\mathbf{R_{X_{\rm d}}D^4R_X}\left(N_0\mathbf{I}_K+\mathbf{D^2R_X}\right)^{-1}\}$. Then, from the Schur concavity of the throughput function, the optimal $\bm{\lambda}$ should be the smallest vector that majorizes $\text{vdiag}\{\mathbf{R_{X_{\rm d}}D^4R_X}\left(N_0\mathbf{I}_K+\mathbf{D^2R_X}\right)^{-1}\}$. Noting the equivalence between problems (\ref{eq:optprob_5}) and (\ref{eq:optprob_10}), we see that the optimal $\mathbf{X}_{\rm p}$ to (\ref{eq:optprob_5}) satisfies the two conditions in the second half of the theorem.

To complete the proof of Theorem \ref{thm:Optimal_pilots_content}, we still need to show that the above pilot design of $\mathbf{X}_{\rm p}$ also minimizes the equivalent noise power $\sigma_v^2$ in (\ref{sigma_v}) (and hence indeed maximizes the throughput). From (\ref{sigma_v}), we see that minimizing $\sigma_v^2$ is equivalent to maximizing
\begin{eqnarray}\label{eqq1}
\mbox{tr} (\mathbf{\widetilde{X}}\mathbf{\widetilde{X}}^\dagger\mathbf{D}^2\mathbf{R}_{\mathbf{X}_{\rm d}})= \mbox{tr} (\mathbf{R}^\frac{1}{2}_{\mathbf{X}_{\rm d}}\mathbf{D\widetilde{X}}\mathbf{\widetilde{X}}^\dagger\mathbf{DR}^\frac{1}{2}_{\mathbf{X}_{\rm d}})
\end{eqnarray}
where $\mathbf{\widetilde{X}}$ is defined in (\ref{eq:X_tilde}). Clearly, $\mbox{tr} (\mathbf{R}^\frac{1}{2}_{\mathbf{X}_{\rm d}}\mathbf{D\widetilde{X}}\mathbf{\widetilde{X}}^\dagger\mathbf{DR}^\frac{1}{2}_{\mathbf{X}_{\rm d}})$ is maximized when the equality in 
(\ref{eq:optprob_52_b}) holds. This agrees with the optimality condition (\ref{eq5}) for the case of $K\leq \alpha T$; it also agrees with condition 1 in Theorem \ref{thm:Optimal_pilots_content} for the case of $K> \alpha T$.\footnote{An explicit construction of $\mathbf{X}_{\rm p}$ satisfying conditions 1 and 2 is presented in Remarks \ref{rmk1} and \ref{rmk2}.} This concludes the proof.

\section{Backgrounds of the Majorization Theory}\label{appendix A}
In this appendix, we introduce some background knowledge of the majorization theory \cite{Majortheory} used in this paper. For $\mathbf{x} = (x_1, \cdots, x_n)$ and $\mathbf{y} = (y_1, \cdots, y_n)$, $\mathbf{x} \succ \mathbf{y}$ if 
\begin{subequations}
\begin{eqnarray}
	\sum_{i=1}^{k}x_{[i]}\!\!&\leq &\!\!\sum_{i=1}^{k}y_{[i]},\ \ \ k = 1, \cdots, n-1\\
	\sum_{i=1}^{n}x_{[i]} \!\!&=&\!\! \sum_{i=1}^{n}y_{[i]}
\end{eqnarray}
\end{subequations}
where $x_{[1]}\leq\cdots\leq x_{[n]}$ and $y_{[1]}\leq\cdots\leq y_{[n]}$ are increasing rearrangements of $\mathbf{x}$ and $\mathbf{y}$, respectively. Let $m$ be an integer satisfying $0\leq m \leq n$. Given a non-negative vector $\mathbf{y} = (y_1, \cdots, y_n)$, the minimum vector $\mathbf{x}^*$ with $m$ zeros that majorizes $\mathbf{y}$ is defined as
\begin{subequations}\label{eqq10}
\begin{eqnarray}
	x_{[1]}^* &=& \cdots = x_{[t]}^* = 0\\
	x_{[t+1]}^* &=&	\cdots = x_{[k]}^* = \frac{1}{k-m}\sum_{i = 1}^{k}y_{[i]}\label{eqq10_b}\\
	x_{[k+1]}^* &=& y_{[k+1]}\\
	&\vdots &\nonumber\\
	x_{[n]}^* &=& y_{[n]},
\end{eqnarray}
\end{subequations}
where $k$ is the smallest number in the index set 
\begin{equation}\label{eqq11xx}
\mathcal{S} = \left\{j \left\vert{\sum_{i = 1}^{j}y_{[i]} \leq (j-m)y_{[j+1]}, m \leq j < n}\right.\right\}\cup\{n\}.	
\end{equation}
In the following two lemmas, we show that $\mathbf{x}^*$ is indeed the \emph{minimum} vector that majorizes $\mathbf{y}$.
\begin{lem}\label{lemma2}
$\mathbf{x}^* \succ \mathbf{y}$.
\end{lem}
\begin{IEEEproof}
We first note that $\mathbf{x}^*$ defined in (\ref{eqq10}), together with $\mathcal{S}$ in (\ref{eqq11xx}), is already arranged in an increasing order. Further, it is clear that
\begin{eqnarray}
	\sum_{i=1}^{k}x_{[i]}^*&\leq& \sum_{i=1}^{k}y_{[i]},\ k = 1,\cdots, n-1\nonumber\\
	\sum_{i=1}^{n}x_{[i]}^* &=& \sum_{i=1}^{n}y_{[i]}.\nonumber
\end{eqnarray}
Therefore, we obtain by definition that $\mathbf{x}^* \succ \mathbf{y}$.
\end{IEEEproof}
\begin{lem}\label{lemma3}
	Let $\mathbf{x}$ be an $n$-by-$1$ nonnegative vector containing $m$ zeroes. Then, $\mathbf{x} \succ \mathbf{y}$ implies $\mathbf{x}\succ\mathbf{x}^*$.  
\end{lem}
\begin{IEEEproof}
From $\mathbf{x}\succ \mathbf{y}$ and (\ref{eqq10}), we obtain
\begin{subequations}\label{eqq11}
\begin{eqnarray}
\sum_{i=1}^{j}x_{[i]}\!\!\!& = &\!\! \sum_{i=1}^{j}x^*_{[i]} \ =\ 0,\ j = 1,\cdots, t\\
\sum_{i=1}^{j}x_{[i]}\!\!\!&\leq &\!\! \sum_{i=1}^{j}y_{[i]} = \sum_{i=1}^{j}x^*_{[i]},\ j = k,\cdots, n-1.
\end{eqnarray}
\end{subequations}
Further, we obtain
\begin{eqnarray}
\frac{1}{j-m}\sum_{i=m+1}^{j}x_{[i]}\leq \frac{1}{k-m}\sum_{i=m+1}^{k}x_{[i]}\leq \frac{1}{k-m}\sum_{i=m+1}^{k}y_{[i]}	,\nonumber\\ 
j = m+1,\cdots, k
\end{eqnarray}
where the first inequality follows from the fact that $x_{[m+1]},x_{[m+2]},\cdots, x_{[k]}$ are in an increasing order, and the second inequality follows from $\mathbf{x}\succ\mathbf{y}$. Together with (\ref{eqq10_b}), we have
\begin{eqnarray}\label{eqq12}
	\sum_{i=m+1}^{j}x_{[i]}\leq\frac{j-m}{k-m}\sum_{i=m+1}^{k}y_{[i]} = \!\!\sum_{i=m+1}^{j}x_{[i]}^*,\ j = m+1,\cdots, k.	
\end{eqnarray} 
The proof concludes by combining (\ref{eqq11}) and (\ref{eqq12}).
\end{IEEEproof}

\section{Proof of Theorem \ref{thm7}}\label{A3}
Let $\widetilde{\mathbf{G}}$ be the submatrix of $\mathbf{G}$ that consists of the first $K$ columns of $\mathbf{G}$. Then, $R = (1-\alpha)\log\left|\mathbf{I}_N+\widetilde{\mathbf{G}}{\rm diag}\{\frac{\lambda_1}{\sigma_v^2},\cdots, \frac{\lambda_K}{\sigma_v^2}\}\widetilde{\mathbf{G}}^\dagger \right|$. As $\rho_0 \rightarrow \infty$, $\log\left|\mathbf{I}_N+\widetilde{\mathbf{G}}{\rm diag}\{\frac{\lambda_1}{\sigma_v^2},\cdots, \frac{\lambda_K}{\sigma_v^2}\}\widetilde{\mathbf{G}}^\dagger \right|$ can be approximated by $\log\left(\prod_{k=1}^K \lambda_k\right) - K\log\sigma_v^2 + \log\left|\widetilde{\mathbf{G}}^\dagger\widetilde{\mathbf{G}} \right|$. Then, the throughput can be written as 
\begin{eqnarray}\label{eqqqx1}
	R = (1-\alpha)\sum_{k=1}^{K} \log \frac{(1-\alpha\gamma_k)\rho_0d_k^2}{\frac{1}{\alpha T}\sum_{k=1}^{K}(\frac{1}{\gamma_k}-\alpha)+ 1-\alpha} \nonumber\\
	+ (1-\alpha)\log\left|\widetilde{\mathbf{G}}^\dagger\widetilde{\mathbf{G}} \right|.
\end{eqnarray} 
Then, the optimization problem reduces to
\begin{subequations} \label{eq:optprob_3011}
\begin{align}
\underset{\{\gamma_k\}}{\mbox{maximize\,\,\,\,}} &  R \ {\rm in} \ (\ref{eqqqx1}) \label{eq:optprob_3011_a}\\
\mbox{subject to\,\,\,\,} & 0\leq \gamma_k\leq \frac{1}{\alpha},\ k \in \mathcal{I}_K. \label{eq:optprob_3011_b}
\end{align}
\end{subequations}
The solution to problem (\ref{eq:optprob_3011}) is described as follows. Note that $\log (1-\alpha\gamma_k)$ is concave in $\gamma_k$, and that $\log (\frac{1}{\alpha T}\sum_{k=1}^{K}(\frac{1}{\gamma_k}-\alpha)+1-\alpha)$ is convex in $\{\gamma_k\}$. Thus, $R$ in (\ref{eqqqx1}) is concave in $\{\gamma_k\}$, and so problem (\ref{eq:optprob_3011}) is a convex problem. By solving the KKT conditions, we obtain $\gamma_1 = \cdots = \gamma_{K} = \gamma$ and $\gamma_1^\prime = \cdots = \gamma_{K}^\prime = \gamma^\prime$, which concludes the proof.

\section{Proof of Theorem \ref{thm6}}\label{A4}
From Corollary \ref{corollary 1}, we see that Algorithm \ref{alg2} outputs the optimal $\mathbf{X}_{\rm p}$ in the sense of throughput maximization when $K \leq \alpha T$. The corresponding optimal $\{\gamma_k\}$ and $\{\gamma_k^\prime\}$ at high SNR are given by Theorem \ref{thm7}, which is in agreement with the statement in Theorem \ref{thm6}. Then, it suffices to focus on the case of $K > \alpha T$, i.e., $\mathbf{X}_{\rm p}$ is a tall matrix.

Without loss of generality, we assume that there are $K^\prime$ active users with $K^\prime\leq K$.\footnote{For an inactive user $k$, we have $\gamma_k = \gamma_k^\prime = 0$.} Let $\mathbf{\overline{X}}_{\rm p}\in \mathbb{C}^{K^\prime\times\alpha T}$ be the pilot matrix for those active users, and $\mathbf{\overline{D}} \in \mathbb{C}^{K^\prime\times K^\prime}$ be the corresponding diagonal fading-coefficient matrix. We consider the limiting process of $\mathbf{X}_{\rm p} = \sqrt\rho_0\mathbf{\overline{W}}_{\rm p}$ as $\rho_0 \rightarrow \infty$, where $\mathbf{\overline{W}}_{\rm p}$ is an arbitrary constant invariant to $\rho_0$.

Let the compact SVD of $\mathbf{\overline{D}\overline{W}}_{\rm p}$ be $\mathbf{\overline{D}\overline{W}}_{\rm p} = \mathbf{\overline{U}}_{\rm p}\mathbf{\overline{\Sigma}}_{\rm p} \mathbf{\overline{V}}_{\rm p}^\dagger$, where $\mathbf{\overline{U}}_{\rm p}\in \mathbb{C}^{K^\prime\times r}$ satisfies $\mathbf{\overline{U}}_{\rm p}^{\dagger}\mathbf{\overline{U}}_{\rm p} = \mathbf{I}_{r}$, $\mathbf{\overline{\Sigma}}_{\rm p}\in \mathbb{C}^{r\times r}$ is a diagonal matrix with the diagonal elements being the singular values, $\mathbf{\overline{V}}_{\rm p}\in \mathbb{C}^{\alpha T\times r}$ satisfies $\mathbf{\overline{V}}_{\rm p}^\dagger\mathbf{\overline{V}}_{\rm p} = \mathbf{I}_{r}$, and $r\ (\leq \alpha T)$ is the rank of $\mathbf{\overline{W}}_{\rm p}$. Then, as $\rho_0\rightarrow\infty$, we obtain
\begin{subequations}
\begin{eqnarray}
&&\mathbf{\overline{D}\overline{X}}_{\rm p}(\mathbf{\overline{X}}_{\rm p}^\dagger \mathbf{\overline{D}}^2\mathbf{\overline{X}}_{\rm p}+N_0\mathbf{I}_{\alpha T})^{-1}\mathbf{\overline{X}}_{\rm p}^\dagger \mathbf{\overline{D}}\\
&=& \rho_0\mathbf{\overline{D}\overline{W}}_{\rm p}(\rho_0\mathbf{\overline{W}}_{\rm p}^\dagger \mathbf{\overline{D}}^2\mathbf{\overline{W}}_{\rm p}+N_0\mathbf{I}_{\alpha T})^{-1}\mathbf{\overline{W}}_{\rm p}^\dagger \mathbf{\overline{D}}\\
&=& \rho_0\mathbf{\overline{U}}_{\rm p}\mathbf{\overline{\Sigma}}_{\rm p}(\rho_0\mathbf{\overline{\Sigma}}_{\rm p}^\dagger\mathbf{\overline{\Sigma}}_{\rm p}+N_0\mathbf{I}_r)^{-1}\mathbf{\overline{\Sigma}}_{\rm p}\mathbf{\overline{U}}_{\rm p}^\dagger\\
&\rightarrow & \mathbf{\overline{U}}_{\rm p}\mathbf{\overline{U}}_{\rm p}^\dagger.
\end{eqnarray}
\end{subequations}
Thus, as $\rho_0 \rightarrow \infty$, the equivalent noise power $\sigma_v^2$ in (\ref{sigma_v}) satisfies
\begin{eqnarray}
\sigma_v^2\!\!&\rightarrow &\!\! {\rm tr}\left\{(\mathbf{I}_{K^\prime}-\mathbf{\overline{U}}_{\rm p}\mathbf{\overline{U}}_{\rm p}^\dagger)\mathbf{\overline{D}}^2\mathbf{\overline{R}}_{\mathbf{X}_d}\right\}+N_0 \label{eqqq2}
\end{eqnarray}
where $\mathbf{\overline{R}}_{\mathbf{X}_d} \in \mathbb{C}^{K^\prime\times K^\prime}$ is the diagonal matrix obtained by deleting the rows and columns of $\mathbf{R}_{\mathbf{X}_d}$ corresponding to inactive users. Note that $\mathbf{I}_{K^\prime}-\mathbf{\overline{U}}_{\rm p}\mathbf{\overline{U}}_{\rm p}^\dagger \neq \mathbf{0}$ for $K^\prime > \alpha T$, i.e., $\sigma_v^2$ in (\ref{eqqq2}) is unbounded when $K^\prime > \alpha T$. This implies that the optimal number of active users for throughput maximization cannot exceed $\alpha T$. Moreover, from Theorem \ref{thm7}, to maximize throughput in the case of $K^\prime \leq \alpha T$, all the $K^\prime$ users should be active. Therefore, the optimal number of active users for throughput maximization is given by $K^\prime = \alpha T$.

We now discuss how to select the $\alpha T$ active users. Without loss of generality, we assume that users $1, \cdots, \alpha T$ are the selected active users. This implies that $\gamma_{\alpha T+1}=\cdots =\gamma_{K} = \gamma_{\alpha T+1}^\prime=\cdots =\gamma_{K}^\prime = 0$. Then, as $\rho_0 \rightarrow \infty$, the throughput in (\ref{eqxqq1}) can be approximated by $\log\prod_{k=1}^{\alpha T} \left(\frac{\lambda_k}{\sigma_v^2}\right) + \log\left|\widetilde{\mathbf{G}}^\dagger\widetilde{\mathbf{G}} \right|$, where $\sigma_v^2$ is given by (\ref{eqq2}), $\lambda_k$ is given by (\ref{eqi1}) for $k = 1,\cdots, \alpha T$, and $\widetilde{\mathbf{G}} \in \mathbb{C}^{N\times\alpha T}$ consists of the first $\alpha T$ columns of $\mathbf{G}$. Further note that, to maximize $\prod_{k=1}^{\alpha T} \left(\frac{\lambda_k}{\sigma_v^2}\right)$ at high SNR, $\alpha\gamma_k+(1-\alpha)\gamma^\prime_k = 1$ for $k = 1, \cdots, \alpha T$. Then, the throughput is given by (\ref{eqqqx1}) with $K$ replaced by $\alpha T$. From (\ref{eqqqx1}), we see that the optimal choice to maximize the throughput is to select users with the maximum $d_k$ values. Finally, the optimal $\{\gamma_k\}$ for the active users are given by Theorem \ref{thm7}, which concludes the proof.

\section{Proof of Theorem \ref{thm3}} \label{A5}
Following the proof of Theorem \ref{thm:Optimal_pilots_content}, we first ignore the effect of $\sigma_v^2$ in throughput maximization. We will discuss the minimization of $\sigma_v^2$ (so as to maximize the throughput) at the end of the proof.

With $d_1 = d_2 = \cdots = d_K = d$ and $\gamma_1 = \gamma_2 = \cdots = \gamma_K = \gamma$, we can rewrite (\ref{eq:R_X_tilde}) as
\begin{equation}\label{eq0x}
R = (1-\alpha)\mathbb{E}\left[\log\left|\mathbf{I}_N+\tfrac{\gamma^\prime d^2P_0}{\sigma_v^2}\mathbf{G}\mathbf{\widetilde{X}}^\dagger\mathbf{\widetilde{X}}\mathbf{G}^\dagger \right| \right]
\end{equation} 
where $\gamma^\prime = \frac{1-\alpha \gamma}{1-\alpha}$, and 
\begin{eqnarray}
\mathbf{\widetilde{X}} &=& d\mathbf{X}_{\rm p}(d^2\mathbf{X}_{\rm p}^\dagger\mathbf{X}_{\rm p}+N_0\mathbf{I}_{\alpha T})^{-\frac{1}{2}}\\
&=& \mathbf{U}_{\rm p}\mathbf{\Sigma}_{\rm p}(\mathbf{\Sigma}_{\rm p}^\dagger\mathbf{\Sigma}_{\rm p}+N_0\mathbf{I}_{\alpha T})^{-\frac{1}{2}}\mathbf{V}_{\rm p}^\dagger
\end{eqnarray}
where the last step follows from the singular value decomposition: $d\mathbf{X}_{\rm p} = \mathbf{U}_{\rm p}\mathbf{\Sigma}_{\rm p}\mathbf{V}_{\rm p}^\dagger$. Then, as $\mathbf{G}\mathbf{V}_{\rm p}$ has the same distribution as $\mathbf{G}$ does, we can further rewrite (\ref{eq0x}) as
\begin{equation}\label{eq1x}
R = (1-\alpha)\mathbb{E}\left[\log\left|\mathbf{I}_N+\tfrac{1}{\sigma_v^2}\mathbf{G}\mathbf{\Lambda}\mathbf{G}^\dagger \right| \right]
\end{equation}
where $\mathbf{\Lambda} \in \mathbb{C}^{\alpha T\times \alpha T}$ is given by
\begin{eqnarray}
\mathbf{\Lambda} \!\!\!\!&=&\!\!\!\! \gamma^\prime d^2P_0(\mathbf{\Sigma}_{\rm p}^\dagger\mathbf{\Sigma}_{\rm p}\!+\!N_0\mathbf{I}_{\alpha T})^{\frac{1}{2}}\mathbf{\Sigma}_{\rm p}^\dagger\mathbf{\Sigma}_{\rm p}(\mathbf{\Sigma}_{\rm p}^\dagger\mathbf{\Sigma}_{\rm p}\!+\!N_0\mathbf{I}_{\alpha T})^{\frac{1}{2}}\nonumber\\
\!\!\!\!&=&\!\!\!\! {\rm diag}\left\{\frac{\gamma^\prime d^2P_0\sigma_1^2}{\sigma_1^2\!+\!N_0},\cdots, \frac{\gamma^\prime d^2P_0\sigma_{\min\{K, \alpha T\}}^2}{\sigma_{\min\{K, \alpha T\}}^2\!+\!N_0}, 0,\cdots, 0\right\}.\nonumber
\end{eqnarray}
Also, the eigenvalue vector of $d^2\mathbf{X}_{\rm p}\mathbf{X}_{\rm p}^\dagger$ can be expressed as $\mathbf{x} = (\sigma_1^2,\cdots,\sigma_{\min\{K, \alpha T\}}^2, 0, \cdots, 0) \in \mathbb{C}^{K}$, where $\sigma_i$ is the $i$th diagonal element of $\mathbf{\Sigma}$. From Theorem 4.3.26 in \cite{horn}, we have $\mathbf{x}\succ {\rm vdiag}(d^2\mathbf{X}_{\rm p}\mathbf{X}_{\rm p}^\dagger) = \alpha\gamma d^2P_0T\mathbf{1}$, where $\mathbf{1}$ is an all-one vector of an appropriate size. Then, problem (\ref{eq:optprob_1}) can be recast as
\begin{subequations} \label{eq:optprobb_1}
\begin{align}
\underset{\mathbf{X}_{\rm p}}{\mbox{maximize\,\,\,\,}} &  R \ {\rm in}\ (\ref{eq1x}) \label{eq:optprobb_1_a}\\
\mbox{subject to\,\,\,\,} & \mathbf{x}\succ \alpha\gamma d^2P_0T\mathbf{1}.  \label{eq:optprobb_1_b}
\end{align}
\end{subequations}
Note that $R$ in (\ref{eqxqq1}) is symmetric and concave with respect to $\sigma_1^2, \cdots, \sigma_{\min\{K,\alpha T\}}^2$. From Proposition C.2 in \cite{Majortheory}, $R$ in (\ref{eqxqq1}) is Schur concave. Thus, the optimal $\bm{x}$ should be the smallest vector that majorizes $\alpha\gamma d^2P_0T\mathbf{1}$. More specifically, for the case of $K\leq \alpha T$, zero padding in $\mathbf{x}$ is not necessary. Thus, the optimal $\mathbf{x}$ is simply taken as $\mathbf{x} = \alpha\gamma d^2P_0T\mathbf{1}$. This implies that $\mathbf{X}_{\rm p}\mathbf{X}_{\rm p}^\dagger = \alpha\gamma P_0T\mathbf{I}_{K}$. For the case of $K> \alpha T$, as the rank of $\mathbf{\Lambda}$ is at most $\alpha T$, we see that $\bm{x}$ is padded with at least $K-\alpha T$ zeros. Then, the optimal $\mathbf{x}$ is given by $x_1 = \cdots = x_{\alpha T} = \gamma d^2 K P_0$ and $x_{\alpha T+1} = \cdots = x_{K} = 0$. This implies that (\ref{eq:XpX}) holds.

To complete the proof, we still need to show that the above pilot design of $\mathbf{X}_{\rm p}$ minimizes the equivalent noise power $\sigma_v^2$ in (\ref{sigma_v}). The detailed argument is very similar to  the corresponding part of the proof of Theorem \ref{thm:Optimal_pilots_content} in Appendix \ref{A2}. We omit the details for brevity. This concludes the proof of Theorem \ref{thm3}.

\section{Proof of Theorem \ref{lem:KleqalphaT}}\label{A6}
We first show that for any given $\alpha$, the optimal $K$ always satisfies $K \geq \alpha T$. To see this, we need the following two facts: 1) $\tau K$ is monotonically increasing in $K$ when  $K \leq \alpha T$; 2)  ${\mathbb{E}}\left[  \log\left|\mathbf{I}_{N}\!+\frac{\theta}{K}\widetilde{\mathbf{G}}\widetilde{\mathbf{G}}^{\dagger}\right| \right]$ is monotonically increasing in $K$ for $K \leq \alpha T$, where $\theta$ is an arbitrary positive constant. Based on these two facts, we see that the throughput in (\ref{eq11}) is monotonically increasing in $K$ for $K\leq \alpha T$. This implies $K^{\rm opt} \geq \alpha T$. In what follows, we prove these two facts.

We first consider Fact 1 as follows. From Theorem \ref{thm:optimal_power}, the optimal $\gamma$ and $\gamma^\prime$ for $K\leq\alpha T$ satisfy the following: $\gamma^{-1} = \alpha(1+\sqrt{1-\mu_1})$ and ${\gamma^\prime}^{-1} = (1-\alpha)\left(\frac{1}{\sqrt{1-\mu_1}}+1\right)$ and ${\gamma^{-1}}{\gamma^\prime}^{-1} = \alpha {\gamma^\prime}^{-1}+(1-\alpha)\gamma^{-1}$, where $\mu_1 = \tfrac{\rho_0d^2(K-(1-\alpha )T )}{1-\alpha+\rho_0d^2K}$. Then 
\begin{eqnarray}
\tau K &=& \frac{\alpha\rho_0^2 d^4KT}{\rho_0d^2(\gamma^{-1} K+\alpha\gamma^{\prime-1} T)+\gamma^{-1}\gamma^{\prime-1}}\nonumber\\
&=& \frac{\rho_0^2d^4KT}{(1{-}\alpha)\left(\sqrt{1{+}\rho_0d^2 T}{+}\sqrt{1{+}\tfrac{\rho_0d^2K}{1-\alpha}}\right)^2}
\label{eq:g1}
\end{eqnarray} 
and
\begin{equation} \small
\frac{{\partial}(\tau K)}{{\partial}K} = \frac{\rho_0^2d^4T\left(\!\sqrt{1\!+\!\rho_0 T}\!+\!\frac{1}{\sqrt{1+ \frac{\rho_0d^2K}{1-\alpha}}}\right)}{(1-\alpha)\left(\sqrt{1+\rho_0d^2T}+\sqrt{1+ \frac{\rho_0d^2K}{1-\alpha}}\right)^3}>0.
\end{equation}
Therefore, $\tau K$  is monotonically increasing in $K$ for $K\leq \alpha T$.

We now consider Fact 2. Denote
\begin{equation}
g(x_1,\cdots, x_K) = {\mathbb{E}}\left[  \log\left|\mathbf{I}_{N}\!+\widetilde{\mathbf{G}}{\rm diag}\{x_1, \cdots, x_K\}\widetilde{\mathbf{G}}^{\dagger}\right| \right].
\end{equation}
Clearly, $g(x_1,\cdots, x_K)$ is symmetric and concave with respect to $x_1,\cdots, x_K$. Thus, from Proposition C.2 in \cite{Majortheory}, $g(x_1,\cdots, x_K)$ is Schur-concave in $x_1,\cdots, x_K$. Let $\mathbf{\widetilde{G}}^\prime\in\mathbb{C}^{N\times\min\{K^\prime, \alpha T\}}$ with the elements independently drawn from $\mathcal{CN}(0,1)$. Then, we obtain
\begin{eqnarray}
{\mathbb{E}}\left[  \log\left|\mathbf{I}_{N}\!+\frac{\theta}{K}\widetilde{\mathbf{G}}\widetilde{\mathbf{G}}^{\dagger}\right| \right] &=& g(\theta/K,\cdots,\theta/K)\nonumber\\
&\geq& g(\theta/K^\prime, \cdots, \theta/K^\prime, 0,\cdots, 0)\nonumber \\
&=& {\mathbb{E}}\left[ \log\left|\mathbf{I}_{N}\!+\frac{\theta}{K^\prime}\widetilde{\mathbf{G}}^\prime \widetilde{\mathbf{G}}^{\prime\dagger}\right| \right]\nonumber
\end{eqnarray}
where the inequality holds from the Schur-concavity of the $g$-function and the fact that 
\begin{eqnarray}
(\underbrace{\theta/K^\prime, \cdots, \theta/K^\prime}_{K^\prime \ \text{times}}, \underbrace{0,\cdots, 0}_{K-K^\prime\ {\rm times}}) \succ (\theta/K, \cdots, \theta/K)
\end{eqnarray}
for any integer $K^\prime < K$. Therefore, Fact 2 holds.

What remains is to determine the optimal $K$ to maximize $R$ in (\ref{eq11}) for $K\geq \alpha T$. In this case, $R$ in (\ref{eq11}) depends on $K$ only through $\tau$. Thus, it suffices to find the optimal $K$ to maximize $\tau$. 

To proceed, we note that for $K\geq \alpha T$, $\gamma^{-1} = \alpha(1+\sqrt{1-\mu_2})$ and ${\gamma^\prime}^{-1} = (1-\alpha)\left(\frac{1}{\sqrt{1-\mu_2}}+1\right)$ and ${\gamma^{-1}}{\gamma^\prime}^{-1} = \alpha {\gamma^\prime}^{-1}+(1-\alpha)\gamma^{-1}$, where $\mu_2 = \frac{(1-\alpha)(\rho_0d^2K+\alpha)}{\alpha(\rho_0d^2K+1-\alpha)}$. Then
\begin{eqnarray}
\tau &=& \frac{\rho_0^2 d^4KN}{\rho_0^2d^4K(K-\alpha T)+\rho_0d^2K(\gamma^{-1} +\gamma^{\prime-1} )+\gamma^{-1}\gamma^{\prime-1}}\nonumber\\
&=& \frac{\rho_0^2 d^4KN}{\rho_0^2d^4K(K-\alpha T)+t(K)^2}
\label{eq:g2}
\end{eqnarray} 
where
\begin{equation}
t(K) = \sqrt{(1-\alpha)(\rho_0d^2K+\alpha)}+\sqrt{\alpha(\rho_0d^2K+1-\alpha)}.
\end{equation}
It is not difficult to see that the above $\tau$, as a function of $K$, has exactly one extremal point (which is the maximum) in the range of $K \geq \alpha T$. Taking derivative of $\tau$ with respective to $K$ and setting it to zero, we see that the optimal $K$ satisfies $f(\rho_0d^2K) = 0$ with $f(x)$ defined in (\ref{eq12}). This concludes the proof of Theorem \ref{lem:KleqalphaT}.

\ifCLASSOPTIONcaptionsoff
\newpage
\fi

%

\end{document}